\documentclass[12pt]{article}
\pdfoutput=1
\usepackage{latexsym}
\usepackage{cite,epsfig,amssymb,euscript,slashed}
\usepackage[normalem]{ulem} %defines \sout{word} which strikes out {word}
\usepackage{amsmath}
\usepackage{float}
\usepackage{array,calc,epsfig}
\usepackage[linktocpage]{hyperref}
\usepackage{bbm}
\usepackage{braket}
\usepackage{xcolor}
\usepackage{fancybox}

\font\mybb=msbm10 at 10pt
\def\bb#1{\hbox{\mybb#1}}

\def\be{\begin{equation}}
\def\ee{\end{equation}}
\def\bseq{\begin{subequations}}
\def\eseq{\end{subequations}}

\def\bea{\begin{eqnarray}}
\def\eea{\end{eqnarray}}

\def\bseq{\begin{subequations}}
\def\eseq{\end{subequations}}

\numberwithin{equation}{section} %%
\usepackage[]{graphicx}

\def\d {{\rm d}}

\def\calc         {{\cal C}}

\def\calh         {{\cal H}}

\def\call         {{\cal L}}
\def\calm         {{\cal M}}
\def\caln         {{\cal N}}

\def\calw         {{\cal W}}

\def\del          {\partial}

\def\ii           {{\rm i}}

\def\tr           {\mathop{\rm Tr}}
\def\Re           {{\rm Re\hskip0.1em}}
\def\Im           {{\rm Im\hskip0.1em}}

\def\sqr#1#2{{\vcenter{\vbox{\hrule height.#2pt
 \hbox{\vrule width.#2pt height#1pt \kern#1pt \vrule width.#2pt}\hrule
 height.#2pt}}}}

\def\g{\gamma}

\def\g{\gamma}

\def\d{\text{d}}

\def\slashchar#1{\setbox0=\hbox{$#1$}           % set a box for #1
\dimen0=\wd0                                 % and get its size
\setbox1=\hbox{/} \dimen1=\wd1               % get siste of /
\ifdim\dimen0>\dimen1                        % #1 is bigger
\rlap{\hbox to \dimen0{\hfil/\hfil}}      % so center / in box
#1                                        % and print #1
\else                                        % / is bigger
\rlap{\hbox to \dimen1{\hfil$#1$\hfil}}   % so center #1
/                                         % and print /
\fi}

\begin{document}
\font\cmss=cmss10 \font\cmsss=cmss10 at 7pt

\title{
%\vspace{-2.0cm}\begin{flushright}{\scriptsize DFPD-2017/TH/10}
%\\  \scriptsize  preprint2}
%\end{flushright}
\hfill\\
\hfill\\
Supermembranes and domain walls \\
in $\mathcal N=1$, $D=4$ SYM
\\[0.5cm]
}

\author{Igor Bandos${}^{a,b}$\footnote{e-mail: {\tt  igor.bandos@ehu.eus }}, Stefano Lanza${}^{c,d}$\footnote{e-mail: {\tt  stefano.lanza@pd.infn.it }} and Dmitri Sorokin${}^{d,c}$\footnote{e-mail: {\tt  dmitri.sorokin@pd.infn.it }}}

\date{}

\maketitle

\vspace{-1.5cm}

\begin{center}

\vspace{0.5cm}\textit{\small
${}^a$ Department of
	Theoretical Physics, University of the Basque Country UPV/EHU,\\
	% Barrio Sarriena s/n, 48940 Leioa, Spain
	P.O. Box 644, 48080 Bilbao
	\\ ${}^b$ IKERBASQUE, Basque Foundation for Science, 48011 Bilbao, Spain}

\vspace{0.5cm}
\textit{\small ${}^c$ Dipartimento di Fisica e Astronomia ``Galileo Galilei",  Universit\`a degli Studi di Padova \\
${}^d$ I.N.F.N. Sezione di Padova, Via F. Marzolo 8, 35131 Padova, Italy}
\end{center}

\vspace{5pt}

\abstract{We construct a manifestly supersymmetric and kappa-symmetry invariant worldvolume action  describing the coupling of a dynamical membrane to an  $\mathcal N=1$, $D=4$ $SU(N)$ super-Yang-Mills multiplet. Worldvolume scalar fields in this action are a Goldstone and a Goldstino associated with spontaneous breaking, by the membrane, of half of $\mathcal N=1$, $D=4$ supersymmetry.  When the Goldstone fields are set to zero, the model reduces to an $\mathcal N=1$, $d=3$ $SU(N)$ Chern-Simons theory induced by the SYM coupling. We show that, when the membrane couples to the Veneziano-Yankielowicz (VY) effective theory of the  $\mathcal N=1$ SYM, it sources VY bulk field equations, separates two distinct SYM vacua and provides the missing contribution to the tension of BPS saturated domain-wall configurations, for which the membrane serves as a core. As a result, we obtain explicit BPS domain-wall solutions in the Veneziano-Yankielowicz theory.
We also briefly discuss a supersymmetric system of an open membrane having a string attached to its boundary and coupled  to a massive extension of the Veneziano-Yankielowicz model.
}
\noindent

%\noindent {\em Possible comment ............  }

\thispagestyle{empty}

%\vfill
%\vskip 5.mm
%\hrule width 5.cm
%\vskip 2.mm
%{\scriptsize
%\noindent e-mails: {\tt sorokin@pd.infn.it
%}}

\newpage

\setcounter{footnote}{0}

\tableofcontents

\newpage

\section{Introduction}

$\mathcal N=1$ super-Young-Mills (SYM) theories in four space-time dimensions, whose first instances were constructed 45 years ago \cite{Wess:1974jb,Ferrara:1974pu,Salam:1974ig}, still attract great deal of attention, both as a base for phenomenological model building and as quantum field theories with a rich vacuum structure. In particular, an $\mathcal N=1$ SYM theory with a gauge group $SU(N)$ has $N$ degenerate supersymmetric vacua associated with different values of the gluino condensate, as was first conjectured in \cite{Witten:1982df}.\footnote{To be concrete, in this paper we will deal with the unitary gauge groups $G=SU(N)$, though the obtained results are valid for a generic simply connected $G$, in which case the number of SYM vacua is equal to the dual Coxeter number $h(G)$ of the gauge group $G$.} As such, there should exist domain wall configurations of SYM fields which interpolate between spatial regions of the theory with two different vacua and preserve one-half of $\mathcal N=1$, $D=4$ supersymmetry (so called 1/2 BPS domain walls) \cite{Dvali:1996xe}.\footnote{Generic properties of BPS domain walls in $N=1$, $D=4$ supersymmetric theories (and of other extended solitons in various dimensions) were considered earlier in \cite{Townsend:1987yy,Abraham:1990nz}.} The domain walls in pure SYM theories and in super Quantum Chromodynamics (SQCD) (with $F$ flavours of matter in the fundamental representation of the gauge group) have been under an extensive study (see \cite{Bashmakov:2018ghn} for the latest review and developments).

These studies have led to a rather comprehensive understanding of properties of domain walls from various perspectives, including construction of SQCD domain wall solutions \cite{Smilga:1997pf,Smilga:1997cx,Smilga:1997yp} in four-dimensional (Wess-Zumino-like) effective field theories (for review and references see e.g.\cite{Shifman:2009zz},\cite{Bashmakov:2018ghn}) and their dual description as three-dimensional gauge theories on worldvolumes of D-branes originating from compactifications of M/String Theory \cite{Acharya:2001dz},\cite{Bashmakov:2018ghn}. However, still some strokes can be added to make this picture more complete.

One of them is the explicit inclusion into $3d$ worldvolume theory of SYM and SQCD domain walls of a Goldstone sector which makes them dynamical objects (membranes) moving in the four-dimensional bulk. The Goldstone sector should consist of an $\mathcal N=1$, $d=3$ scalar supermultiplet. Its scalar component is the Goldstone of spontaneously broken translations in the direction traverse to the domain wall and its Majorana spinor component is the Goldstino associated with the broken half of $\mathcal N=1$, $D=4$ supersymmetry. The spontaneously broken part of $\mathcal N=1$, $D=4$ supersymmetry transformations is non-linearly realized on the Goldstone supermultiplet.

Another open issue is the explicit construction of domain wall solutions in pure (strongly coupled) SYM theory. Attempts to realize this construction  within the Veneziano-Yankielowiz (VY) effective field theory of $\mathcal N=1$ SYM \cite{Veneziano:1982ah} have been undertaken in \cite{Kovner:1997ca,Kogan:1997dt}. The VY theory is a generalized Wess-Zumino model describing gluino-balls, i.e. a chiral scalar supermultiplet formed by $SU(N)$ singlets of bi-linears of gluinos and their superpartners. It is not an effective theory in the Wilsonian sense, since it does not describe all lightest SYM modes, which should also include e.g. other types of glueballs (see e.g. \cite{Shore:1982kh,Intriligator:1994jr,Intriligator:1995au,Kovner:1997ca,Farrar:1997fn} for the discussion of this issue). However, the VY superpotential exactly captures the vacuum structure of $\mathcal N=1$ SYM theory and one may also tempt to use it for studying BPS domain walls separating two SYM vacua \cite{Kovner:1997ca,Kogan:1997dt}.  The potential obtained by integrating out the auxiliary fields in the VY action (amended in \cite{Kovner:1997im}) has a ``glued" structure with cusps separating each neighbouring pair of the $N$ SYM vacua. The domain wall tension (naively) estimated with the use of the VY superpotential was shown \cite{Kogan:1997dt} to be much smaller than the exact value of the tension of the BPS saturated domain walls. Ref. \cite{Kogan:1997dt} suggested that at the cusp of the potential there should live an object (associated with integrated heavy modes of the theory) whose contribution restores the BPS value of the domain wall tension. To our knowledge, this object has not been identified yet.

In this paper we will show that the solution of the first problem, i.e. the construction of a manifestly supersymmetric and kappa-symmetry invariant action for a dynamical membrane of a charge $k$ coupled to $\mathcal N=1$, $D=4$ SYM, also solves the second problem. Namely, when the membrane couples to the VY effective theory it sources VY field equations, separates two distinct SYM vacua and provides the missing contribution to the tension of the BPS saturated domain-wall configurations. As a result, we obtain explicit BPS domain wall solutions in the VY theory.

We also show that the worldvolume theory of the static membrane, which is obtained when the Goldstone field fluctuations are set to zero, is (in the conventions of \cite{Bashmakov:2018ghn}) an $\mathcal N=1$, $d=3$ $SU(N)_{-k}$ Chern-Simons theory of level $k$, where $0<k<N$ is the membrane charge inducing the transition from the $n$-th  to the $(n+k)$-th SYM vacuum on different sides of the membrane.\footnote{A bosonic membrane coupled to gauge fields via the Chern-Simons term (which is closely related to our supersymmetric construction) was considered in a widely unknown paper \cite{Townsend:1993wy} as an effective description of axionic defects. For even earlier generic constructions of couplings of $p$-branes to Young-Mills fields in various dimension see  \cite{Dixon:1991xz,Dixon:1992qd}.} For $k=1$, the obtained  $SU(N)_{-1}$ Chern-Simons theory is level/rank dual \cite{Hsin:2016blu,Gomis:2017ixy,Bashmakov:2018wts,Bashmakov:2018ghn}  to  Acharya-Vafa (AV) \cite{Acharya:2001dz} $U(1)_N$ worldvolume theory  of the $k=1$ domain wall.
For $k\geq 1$ the Acharya-Vafa  theory is a three-dimensional ${\mathcal N} = 1$ $U(k)_N$ gauge theory with the CS term of level $N$ and an adjoint scalar multiplet.
Upon integrating out the heavy adjoint scalar multiplet, in the assumption that its fermionic mass is negative, one gets the low-energy description of AV theory as a $U(k)_{N-\frac k2,N}$ gauge theory which further reduces (upon integrating out gluini) to a topological $U(k)_{N-k,N}$ CS theory (see \cite{Gomis:2017ixy,Bashmakov:2018wts}, and also \cite{Bashmakov:2018ghn} for a review).
In our case, for $k>1$ the level/rank duality \cite{Hsin:2016blu} maps the $SU(N)_{-k}$ Chern-Simons theory to a $U(k)_{N,N}$ one, which is different from the low-energy description of the AV theory.\footnote{Formally, this $SU(N)_{-k}$ Chern-Simons theory seems to correspond to an infrared limit of a different $\mathcal N=1$, $d=3$ theory with a gauge group $SU(N)_{-k}$ and one adjoint scalar multiplet considered in \cite{Bashmakov:2018wts}.}

This discrepancy may be due to the fact that in our construction we are dealing with a single membrane of a charge $k$ which does not accommodate all the (non-Abelian) fields leaving on the $k$-wall. To incorporate the missing fields, one should regard the membrane of charge $k$ as the center of mass of $k$ coincident branes of charge 1 and excite relative fluctuations of these membranes around the center of mass, by analogy with the theory of a stack of $k$ coincident D-branes. The complete structure of such a non-Abelian Born-Infeld-like action which respects supersymmetry and worldvolume symmetries is yet unknown and its construction is beyond the scope of our paper.
In this respect we would just like to stress that our model (though not capturing all the details of the worldvolume theory) correctly reproduces the BPS saturated tension of the SYM domain walls and the corresponding tensorial central charge (including its phase) in the ${\mathcal N}=1$, $D=4$ supersymmetry algebra. It also makes possible to explicitly construct BPS domain wall configurations in the framework of the Veneziano-Yankielowicz effective theory. In particular, for membranes with relatively small charges $|k|\leq \frac N3$ we find regular half-BPS domain walls (with the complex scalar gluino-ball field
being continuous through the membrane).

In addition, we shall also consider a system of an open membrane with a string attached to its boundary coupled to a massive three-form superfield extension \cite{Burgess:1995kp,Farrar:1997fn} of the Veneziano-Yankielowicz theory, which may be applied to the study of domain-wall junctions.

The paper is organized according to its Table of Contents.
We use the notation and conventions of \cite{Bandos:2018gjp} (i.e. mostly that of  \cite{Wess:1992cp} and \cite{Buchbinder:1995uq}, see also Appendix \ref{conventions}).

\section{Overview of the \texorpdfstring{$\mathcal N=1$, $D=4$ $SU(N)$}{N=1,D=4 SU(N)} SYM}

The $\mathcal N=1$ SYM multiplet consists of a gauge vector field $A_m(x)$ and its fermionic superpartners which, in the two-component Weyl spinor notation, are $\lambda_\alpha(x)$ and its complex conjugate $\bar\lambda_{\dot\alpha}(x)$. The SYM multiplet also contains an auxiliary scalar field $D(x)$ to make the supersymmetry transformations acting on the fields form the closed off-shell superalgebra. All the fields mentioned above take values in the adjoint representation of $SU(N)$ whose indices will be suppressed.

The SYM Lagrangian has the following well-known form
\be\label{SYML}
\begin{split}
\call_{\rm SYM}=&-\frac {\ii} {2g^2}\tr\lambda\sigma^m \nabla_m\bar\lambda + \frac {\ii} {2g^2}\tr\nabla_m \lambda\sigma^m \bar\lambda-\frac 1{4g^2} \tr F_{mn}F^{mn}+\frac 1{2g^2}\tr D^2\\
&  +\frac \vartheta {32\pi^2} \tr(\varepsilon_{mnpl}F^{mn}F^{pl} +4\partial_m (\lambda\sigma^m\bar\lambda))\,,
\end{split}
\ee
where $F_2=\d A +\ii A\wedge A$ is the gauge field strength, $\nabla_m=\partial_m  - \ii A_m$, $g$ is the SYM coupling constant and $\vartheta$ is the angle of the topological term.

In the superfield formalism, the SYM Lagrangian is constructed as an F-term, i.e. as an integral over chiral Grassmann coordinates $\theta^\alpha$
\be\label{SYML1}
\call_{\rm SYM}=\frac\tau{8\pi}\int \d^2\theta \tr {\mathcal W}^\alpha{\mathcal W}_\alpha+{\rm c.c.}\,,
\ee
where
$
\tau=\frac{\ii\vartheta}{2\pi}+\frac{2\pi}{g^2}
$
and $\mathcal W_\alpha(x_L,\theta)$ is the chiral superfield accommodating the SYM field-strength multiplet
\be
\label{Wexp}
\mathcal W_\alpha=-\ii\lambda_\alpha +\theta_\alpha D- \frac \ii 2 F_{mn}\sigma^{mn}{}_\alpha{}^\beta\theta_\beta
+\theta^2\sigma^{m}_{\alpha\dot\beta}
\nabla_m\bar{\lambda}{}^{\dot\beta}.
\ee
The classical $U(1)$ R-symmetry of the SYM action (under chiral rotations $\lambda \to \lambda e^{\ii \varphi}$) is broken by quantum anomaly down to a discrete subgroup $Z_{2N}$. The instanton effects create a gluino condensate \cite{Witten:1982df} whose  value was first computed in \cite{Shifman:1987ia} (see also e.g. \cite{Davies:1999uw})
\be\label{gcond}
\braket{\lambda\lambda}\equiv \braket{\tr\lambda^\alpha\lambda_\alpha}\propto \Lambda^3e^{\frac{2\pi\ii n}N}, \quad n=0,1,\ldots, N-1,
\ee
where $\Lambda^3$ is a SYM dynamical scale.

The parameter $n$ labels $N$ degenerate supersymmetric vacua of the SYM theory related by $Z_N$ symmetry. In other words the gluino condensate further breaks the $Z_{2N}$ R-symmetry down to $Z_2$ ($\lambda \to -\lambda$).

\section{Veneziano-Yankielowicz Lagrangian for the \texorpdfstring{\hbox{$\mathcal N=1$}}{N=1} SYM revisited}\label{VYLa}
The gluino condensate and the $N$ vacua of $SU(N)$ SYM are effectively described by the Veneziano-Yankielowicz Lagrangian \cite{Veneziano:1982ah}. The chiral superfield $S$ in this Lagrangian accomodates a gaugino bi-linear (the \emph{gluino-ball})
and its superpartners. It is a composite chiral scalar superfield made of the trace of the bi-linear of the SYM field-strength \eqref{Wexp}
\be\label{S=WW}
S=\tr {\mathcal W}^\alpha {\mathcal W}_\alpha =s+\sqrt 2\theta^\alpha\chi_\alpha+\theta^2F,
\ee
where
\begin{align}
s&=-\tr \lambda^\alpha \lambda_\alpha\,,
\\
\label{chi}
\chi_\alpha&=\sqrt{2} \tr\left(\frac 1 2 \, F_{mn}\sigma^{mn}_{\alpha}{}^\beta\lambda_\beta-\ii\lambda_\alpha D \right),
\end{align}
and
\be\label{F=2}
F=\tr \left(-2\ii\lambda\sigma^m \nabla_m\bar\lambda-\frac 12F_{mn}F^{mn}+D^2 -\frac \ii 4 \varepsilon_{mnpl}F^{mn}F^{pl}\right)\,.
\ee
Note that in the VY theory $s$, $\chi$ and $F$ are regarded as elementary colorless fields.

One can notice \cite{Burgess:1995kp,Binetruy:1996xw} that the superfield $S$ is a special one of the type first introduced in \cite{Gates:1980ay}. It contains, in its $F$-component, the field strength of a (composite) three-form, the latter being the $SU(N)$ Chern-Simons term\footnote{Our complex conjugation rules for the fermions are $(\lambda_1\lambda_2)^*=\bar\lambda_2\bar\lambda_1$, so e.g. $\lambda\sigma^l\bar\lambda$ in \eqref{F4=dCS} is real.}
\bea\label{F4=dCS}
F_4=\d^4x\,\, \Im F&=&-\tr F_2\wedge F_2-\d^4x\, \partial_m (\tr \lambda\sigma^m\bar\lambda)  \qquad \\ &=&- \d\tr \left(A \d A+\frac {2\ii}3 A^3+\frac 1 {3!}
 \d x^k \d x^n\d x^m \epsilon_{mnkl}\tr \lambda\sigma^l\bar\lambda \right)\equiv \d C_3.\nonumber
\eea
Therefore, the complex field $F$ in \eqref{S=WW} has the following form
\be\label{F}
F=\hat D+\ii\partial_mC^m,
\ee
where $\hat D$ is a scalar field and $C^m$ is the Hodge dual of the three-form $C_3$ (see \eqref{HD} for the definition).

Chiral superfields containing field strengths of 3-form fields among their components were introduced in \cite{Gates:1980ay}. In the case of the single three-form superfield like $S$, the chirality constraint $\bar D_{\dot\alpha}S=0$ has the following general solution
\be\label{S=D2U}
S=-\frac 14 \bar D_{\dot\alpha}\bar D^{\dot\alpha} U,
\ee
where $U$ is a real superfield prepotential, and $\bar D_{\dot\alpha}$ and $D_\alpha$ are super-covariant spinor derivatives.\footnote{One should not confuse $U$ with the real SYM prepotential $V$ appearing in the definition of the field strength $\calw_\alpha = - \frac18 \bar D^2 (e^{-2V} D_\alpha e^{2V})$.} The requirement that $U$ should be real rather than complex (which would be the case of a generic chiral field) is connected with the fact that the real $U$ contains the real one-form  $C_1$ dual to $C_3$ among its independent bosonic components
\be\label{U}
\begin{aligned}
U | &= u ,\\
- \frac 18 \bar{\sigma}^{\dot\alpha \alpha}_m [D_\alpha, \bar{D}_{\dot\alpha}] U| &= C_m,\\
\frac1{4} D^2 U| &= -\bar s=\tr \bar\lambda\bar\lambda,\\
\frac{1}{16} D^2 \bar{D}^2 U| &= \hat D + \ii \partial^m C_m \equiv F \, .
\end{aligned}
\ee
We also note that \eqref{S=D2U} is invariant under the gauge transformation
\be\label{UtoU+L}
U'= U+L\,,
\ee
where $L$ is a real linear superfield \be
\label{Lcon}
\bar D^2L=0=D^2L\,.
\ee
Therefore the leading bosonic component of $U$ is a pure gauge.

Below we will show that (in view of \eqref{S=D2U}) the treatment of the superfield $U$ rather than $S$ as the independent superfield  in the Veneziano-Yankielowicz Lagrangian requires the modification of the latter by a certain surface term, whose form is fixed by a consistency of the variation principle with respect to $U$ (see \cite{Groh:2012tf,Farakos:2017jme,Farakos:2017ocw} for  details and references).

The original Veneziano-Yankielowicz Lagrangian is
\be\label{VYL}
\mathcal L_{\rm VY}=\frac 1{16\pi^2\rho} \int \d^2\theta \d^2\bar \theta (S\bar S)^{\frac 13}+\int \d^2\theta\, W(S)+{\rm c.c.}\,,
\ee
in which the VY superpotential is uniquely fixed by  anomalous  superconformal Ward  identities of the SYM theory and has the following form
\be\label{VYsp}
W(S)=\frac N{16\pi^2}S\left(\ln \frac{S}{\Lambda^{3}}-1\right)\,, \qquad W_S:= \partial_SW(S)=\frac N{16\pi^2}\ln \frac{S}{\Lambda^{3}}\,.
\ee
The first term in \eqref{VYL} is the K\"ahler potential
\be\label{K=}
K(S, \bar S)= \frac 1{16\pi^2\rho}  (S\bar S)^{\frac 13}
\ee
whose simplest form is chosen due to the mass dimension 3 of the superfield $S$   and $\rho$ is a dimensionless (a priori arbitrary) positive constant. In general, the kinetic part of the Lagrangian is not fixed by anomalous symmetries and can also include higher order terms \cite{Shore:1982kh}.

One can assume (as in \cite{Kogan:1997dt}) that $\rho$ should scale with $N$ as $\sim \frac 1N$, then the K\"ahler potential term and superpotential would have the same $N$-dependence in the Lagrangian.
However, we prefer to consider a generic $\rho$ since, as we shall see later, its eventual dependence on $N$ affects the characteristic width of the domain walls.

The treatment of the VY Lagrangian as a conventional Wess-Zumino model has encountered a couple of issues \cite{Kovner:1997im}. One of them is that $\tr F_2\wedge F_2$ is the instanton density and  the elimination of this component of the auxiliary field $F_S$ from the
action requires caution. A recipe of how one can take care of this subtlety by modifying the VY superpotential was proposed in \cite{Kovner:1997im}. The fact that the term in question is actually the field strength of the (Chern-Simons) three-form, i.e. that the superfield $S$ is special (see eqs. \eqref{F4=dCS} and \eqref{S=D2U}) was used in \cite{Burgess:1995kp,Farrar:1997fn} to generalize the VY Lagrangian by terms which make the auxiliary components of $U$ dynamical fields describing additional massive glueball states. This construction was further refined in \cite{Cerdeno:2003us} (see also references therein).

In this paper we would like to elaborate on  the role of the special nature of the superfield $S$ within the original VY Lagrangian.  We will see that by treating $U$ as the independent superfield and  modifying the Lagrangian \eqref{VYL} with an appropriate boundary term allows one to consistently eliminate the auxiliary fields by solving their equations of motion and to get additional contributions to the effective scalar potential of (quantized) numerical integration parameters similar to those introduced in \cite{Kovner:1997im}. This also solves the second issue with the VY Lagrangian whose superpotential is not single-valued: because of the presence of the logarithmic term, it gets shifted by  the (identical)  phase transformation
\be\label{StoS'}
S(x,\theta)\to S'(x,\theta e^{\pi \ii})=e^{2\pi \ii}S(x,\theta)\,, \qquad W(S)  \to W(S){+}\frac{\ii N}{8\pi} S\,.
\ee
 The addition of the boundary term compensates the shift in the superpotential and makes the whole Lagrangian single-valued.

The total space-time derivative term in question has the following form
\bea\label{btVY}
{\mathcal L}_{\rm bd}
&=&-\frac 1{128\pi^2} \left(\int\d^2\theta \bar D^2-\int\d^2\bar\theta D^2\right)\left[\left(\frac 1{12\rho } \bar D^2 \frac{\bar S^\frac 13}{{S^\frac 23}}{+}\ln \frac{\Lambda^{3N}}{S^N}\right)U\right]
+\text{c.c.}
\eea
For a general class of models involving three-form chiral supermultiplets the boundary terms of this kind were derived in \cite{Farakos:2017jme}. Their form is singled out by the requirement that the variation of $U$ (and hence $\delta C_m$) is not restricted on the boundary, while $\delta S|_{\rm bd}=0$ and $\delta F_4|_{\rm bd}=0$. This requirement insures the consistency of the variational principle when dealing with the three-form gauge fields in 4D field theories (see \cite{Groh:2012tf,Farakos:2017jme} for a review of this issue and references).

It is not hard to see that the Lagrangian
\be\label{VY+bd}
\mathcal L=\mathcal L_{\rm VY}+\mathcal L_{\rm bd}
\ee
is invariant under the phase transformation \eqref{StoS'}. Actually, it is invariant under a generic $U(1)$  R-symmetry rotation. In order to break this symmetry down to $Z_{2N}$, as it happens in the SYM due to the chiral anomaly, we will require that the term \hbox{$X(S{{, \bar S}})\equiv \frac 1{16\pi^2 }\left(\frac 1{12 \rho } \bar D^2 \frac{\bar S^\frac 13}{{S^\frac 23}}{+}\ln \frac{\Lambda^{3N}}{S^N}\right)$}in the Lagrangian (\ref{btVY})  satisfies the following boundary conditions  $X(S{{, \bar S}})|_{\rm bd}={-}\frac{\ii n}{8\pi}$, where $n=0,1\ldots, (N-1)\,({\rm{mod}}\, N)$ characterizes the asymptotic vacua of the theory. Note that with this choice of the boundary conditions the Lagrangian (\ref{VY+bd}) is gauge invariant under \eqref{UtoU+L}.

Let us now proceed and eliminate the auxiliary field $\hat D$ and the field strength $F_4$ of $C_3$ by solving their equations of motion. To this end let us set the fermions to zero and consider the bosonic part of the Lagrangian \eqref{VY+bd}
\be
\label{LagThreeFormCompVY}
\begin{split}
\call^{\rm bos}_{\rm VY}=  K_{s\bar s}\left({- \partial_m{s}\partial^m{\bar s}}+(\partial_m C^{m})^2+\hat D^2 \right)
 + \left( {W}_s \left(\hat D+\ii\partial_m C^{m}  \right) + \text{c.c.}\right) + \call^{\rm bos}_{\rm bd}\, ,
\end{split}
\ee
with the boundary term
\be
\label{compbound}
\begin{split}
\call^{\rm bos}_{\rm bd} =  - 2\partial_m \left(  C^{m}  K_{s\bar s}  \del_n C^{n}
\right)- \ii\del_m  \left( C^{m} (W_s -\bar{W}_{\bar s})\right) \, ,
\end{split}
\ee
where $S|=-\tr\lambda\lambda\equiv s(x)$, $K(s,\bar s)$ and $W(s)$ are the VY K\"ahler potential and superpotential (at $\theta=\bar\theta=0$), and we have defined $K_{s} \equiv \frac{\del K}{\del s}$, $K_{s\bar s} \equiv \frac{\del^2 K}{\del s \del \bar s}$, $W_s \equiv \frac{\del W}{\del s}$ etc.

Varying the Lagrangian \eqref{LagThreeFormCompVY} with respect to $\hat D$ and $C_m$ we get the following equations of motion
\be\label{D}
K_{s\bar s} \hat D+\Re W_s=0\,,
\ee
\be\label{Cm}
\partial_m(K_{s\bar s} \partial_nC^n-\Im W_s)=0\,.
\ee
From the first of these equations we get the on-shell value of the auxiliary field $\hat D$
\be\label{Dv}
\hat D=-\frac {\Re W_s}{K_{s\bar s}}={9\rho N} (s\bar s)^{\frac 23}\ln{\frac{\Lambda^3}{|s|}}
\ee
and solving the second we get
\be\label{Cmv}
\partial_mC^m=\frac {\Im W_s-\frac{n}{8\pi}}{K_{s\bar s}}=-{9\rho N}(s\bar s)^{\frac 23}\left(2\pi \frac nN-\arg s\right),
\ee
where $n$ is the integration constant parameter compatible with the chosen boundary conditions.

Substituting these expressions back into the Lagrangian \eqref{LagThreeFormCompVY} we get the effective potential for the scalar field
\bea\label{V(S)}
 V(s,\bar s)
 %-\mathcal L^{\rm bos}|_{\partial_ms=0}
 &=&\frac 1{K_{s\bar s}}\left[(\Re W_s)^2+\left(\Im  W_s-\frac n{8\pi}\right)^2\right]\nonumber\\
 &=&{9 \rho N}(s\bar s)^{\frac 23}\left[\ln^2\frac {|s|}{\Lambda^{3}}+\left(2\pi \frac nN-\arg s\right)^2\right]\,.
\eea
When $n=0$ the form of this potential coincides with that of the Veneziano and Yankielowicz, while for  $n=1, 2, \ldots$ it coincides with that of \cite{Kovner:1997ca,Kogan:1997dt}.

The potential is invariant under the simultaneous shifts of the $Z_N$ R-symmetry
\be\label{shift}
 n \to n+k, \qquad \arg s \to \arg s +2\pi \frac kN\,.
\ee
As was argued in \cite{Kovner:1997ca}, the  parameter $n$ should be considered as a discrete variable with respect to which one should take the sum in the functional integral determining the effective action. This makes the potential a continuous function of the phase of $s$ but having cusps at the points in which $n$ changes its values (see \cite{Kovner:1997ca,Kogan:1997dt}  for more details). The potential vanishes when the vevs of $s$ take the values of the gaugino condensate $\braket{s}=-\braket{\lambda\lambda}$ (see eq. \eqref{gcond}).

\section{Supermembranes in \texorpdfstring{$\mathcal N=1$}{N=1} SYM theory}

Now, using the results of \cite{Bandos:2010yy,Bandos:2018gjp}, we will couple a supermembrane to the $\mathcal N=1$ SYM (and its Venezian-Yankielowicz effective action) and study BPS domain wall configurations which it sources in this theory. To our knowledge, the kappa-symmetric action for such a supermembrane has not been considered in the literature yet.\footnote{In $D=4$, the supermembranes have been mainly considered with regard to their couplings to $N=1$, $D=4$ supergravity and chiral matter supermultiplets \cite{Ovrut:1997ur,Huebscher:2009bp,Bandos:2010yy,Bandos:2011fw,Bandos:2012gz,Kuzenko:2017vil,Bandos:2018gjp,Bandos:2019wgy,Bandos:2019khd} which have not included the SYM multiplet.} The addition of the dynamical membrane action to the VY action solves a long-standing issue \cite{Kogan:1997dt} of the discrepancy between the tension of the would-be BPS domain walls calculated in the VY effective theory and the actual tension of the BPS saturated domain walls $T=2|W_{+\infty}-W_{-\infty}|$, where $W_{\pm\infty}$ are the values of the superpotential at the two vacua $\braket{S}_{\pm\infty}$ between which the domain wall is interpolating.

If in a theory we have only a single special chiral three-form superfield like \eqref{S=D2U}, then the most general action describing its coupling to a membrane in flat $\mathcal N=1$, $D=4$ superspace, parametrized by the supercoordinates $z^M=(x^m, \theta^\alpha,\bar\theta^{\dot\alpha})$, has the following form
%%%%%%%%%%%%%

\be\label{susym1}
S_{\rm membrane}=-\frac 1{4\pi}\int_{{\mathcal M}_3}\d^3\xi\sqrt{-\det h_{ij}}\left|kS+c\right| -\frac k{4\pi}\int_{{\mathcal M}_3}{\mathcal C}_3-\left(\frac{\bar c}{4\pi}\int_{{\mathcal M}_3}{\mathcal C}^0_3+c.c.\right),
\ee
where  $c=k_1+ik_2$, and $k$, $k_1$ and $k_2$ are real constant charges characterizing the membrane coupling to a real three-form gauge superfield ${\mathcal C}_3$ and a complex super three-form ${\mathcal C}^0_3$ to be defined below. The  normalization factor $\frac  1{4\pi}$ has been chosen to have the canonical form of the Chern-Simons term in the static membrane action which forces the charge $k$ be quantized  (see Section \ref{msym}).

In the Nambu-Goto part of action \eqref{susym1}  the bulk superfield $S(x,\theta,\bar\theta)$ is evaluated on the membrane worldvolume $z^M=z^M(\xi)$ parametrized by $\xi^i\,(i=0,1,2)$,
\be
h_{ij}(\xi)\equiv \eta_{ab}E^a_i(\xi)E^b_j(\xi),\quad~~\text{with }\quad E^a_i(\xi)\equiv \del_i z^M(\xi)E^a_M(z(\xi)),
\ee
is the induced metric on the membrane worldvolume and
\be\label{sva}
E^a(\xi)\equiv \d z^M(\xi)E_M^a(z(\xi))=\d x^a(\xi)+\ii\theta\sigma^a \d\bar\theta(\xi)-\ii \d\theta\sigma^a \bar\theta(\xi)
\ee
is the worldvolume pull-back of the flat superspace vector supervielbein.

The super three-form $\mathcal C_3$ is constructed in terms of the real prepotential $U$ (see eqs. \eqref{S=D2U} and \eqref{U}) as follows \cite{Gates:1980ay,Binetruy:1996xw}
\be\label{super3form}
\begin{aligned}
\mathcal C_{3}=&\,   {\ii} E^a \wedge \d\theta^\alpha \wedge \d\bar\theta^{\dot\alpha}  \sigma_{a\alpha\dot\alpha}U \\ & - {\frac 14}  E^b\wedge E^a \wedge  \d\theta^\alpha
\sigma_{ab\; \alpha}{}^{\beta}{D}_{\beta}U -{\frac 14}  E^b\wedge E^a \wedge  \d \bar\theta^{\dot\alpha}
\bar\sigma_{ab}{}^{\dot\beta}{}_{\dot\alpha}\bar{D}_{\dot\beta}U
\\&-\frac {1} {48}
  E^c \wedge E^b \wedge E^a \epsilon_{abcd} \,\bar{\sigma}{}^{d\dot{\alpha}\alpha}
  [D_\alpha, \bar{D}_{\dot\alpha}]U
 \, .
\end{aligned}
\ee
Note that the last, purely tensorial, term in \eqref{super3form} coincides, at $\theta=\bar\theta=0$, with the three-form component of $U$ in \eqref{U}.

The associated supersymmetric four-form field strength is
\begin{equation}
\begin{aligned}
\label{bH4=SGZ}
\calh_{4}   &= \d\,{\cal C}_3= \frac18 E^b\wedge E^a \wedge (\d\theta^\alpha \wedge \d\theta_\beta \sigma_{ab\; \alpha}{}^\beta D^2 U
+ \d\bar\theta_{\dot \alpha} \wedge \d\bar\theta^{\dot \beta} \bar \sigma_{ab}{}^{\dot\alpha}{}_{\dot\beta} \bar D^2 U )
 \\
&\quad\, +\frac{1}{48} E^c\wedge E^b\wedge E^a \wedge( d\theta^\alpha \epsilon_{abcd} \sigma^{d}_{\alpha\dot\alpha} \bar{D}^{\dot\alpha} D^2 U
- \d\bar \theta_{\dot\alpha} \epsilon_{abcd} \bar\sigma^{\d\dot\alpha\alpha} D_{\alpha}{ \bar D^2 U})
\\
&\quad\, +\frac{\ii}{8 \times 96} E^{d} \wedge E^c \wedge E^b \wedge E^a \epsilon_{abcd} [D^2, \bar D^2 ] U
\,.
\end{aligned}
\end{equation}
The complex three-form
${\mathcal C}_3^0$
  has the following form\footnote{\label{sugra} Note that, by analogy with \eqref{super3form} one can regard $\mathcal C^0_3$ as the three-form associated to a complex prepotential $\Sigma^0=\theta^2$ satisfying the complex linear constraint $ \bar D^2  \Sigma^0=0$. Then, by analogy with \eqref{S=D2U}, this prepotential gives rise to a trivial ``special chiral superfield" $Z=-\frac 14 \bar D^2 \bar \Sigma=1$. The latter constraint can be interpreted as the gauge fixing condition imposed on a complex three-form conformal compensator superfield $Z$ which fixes the super-Weyl invariance in the $\mathcal N=1$ supergravity coupled to the membrane. From this perspective the three-form \eqref{C03} is the flat superspace remnant  of a complex gauge-three-superform whose dual field strength is traded for the complex auxiliary scalar field of old minimal supergravity \cite{Kuzenko:2017vil},\cite{Bandos:2018gjp}.}
\be\label{C03}
\mathcal C^0_3=\,  {\ii} E^a \wedge \d\theta^\alpha \wedge \d\bar\theta^{\dot\alpha}  \sigma_{a\alpha\dot\alpha}\,\theta^2  -\frac 12  E^b\wedge E^a \wedge  \d\theta^\alpha
\sigma_{ab\; \alpha}{}^{\beta}{\theta}_{\beta} ,
\ee
and its supersymmetry invariant field strength is
\be
\mathcal H^0_4  =- \frac 12 E^b\wedge E^a \wedge  \d\theta^\alpha
\sigma_{ab\; \alpha}{}^{\beta}\d{\theta}_{\beta}.
\ee

\subsection{Local worldvolume symmetries of the membrane action}
By construction, the action \eqref{susym1} is invariant under the worldvolume diffeomorphisms $\xi^i \to f^i(\xi)$ and under the $\kappa$-symmetry transformations
\be\label{kappasymm}
\delta\theta^\alpha=\kappa^\alpha(\xi),\quad \delta \bar\theta^{\dot\alpha}=\bar\kappa^{\dot\alpha}(\xi),\quad \delta x^m=\ii\kappa\sigma^m\bar\theta-\ii\theta\sigma^m\bar\kappa,
\ee
such that
$$
\delta_\kappa z^ME_M^a=0\,.
$$

The local fermionic parameter $\kappa^\alpha(\xi)$ and its complex conjugate $\bar\kappa^{\dot\alpha}(\xi)$ satisfy the following projection condition
\be\label{kappaproj}
\kappa_\alpha=-{\ii}\frac {kS+c}{|kS+c|}\Gamma_{\alpha\dot\alpha}\bar\kappa^{\dot\alpha}  \quad \Leftrightarrow \quad \bar\kappa_{\dot\alpha}=-{\ii}\frac {k\bar S +\bar c}{|kS+c|}\Gamma_{\alpha\dot\alpha}\kappa^{\alpha},
\ee
where
\be\label{kappagamma}
\Gamma_{\alpha\dot\alpha}\equiv \frac{\ii\epsilon^{ijk}}{3!\sqrt{-\det h}}\epsilon_{abcd} E^b_iE^c_j E^d_k\,\sigma^a_{\alpha\dot\alpha}, \qquad \Gamma_{\alpha\dot\alpha}\Gamma^{\dot\alpha\beta}=\delta_\alpha^\beta.
\ee
As is well known, the $\kappa$-symmetry corresponds to half of the bulk supersymmetry preserved by a BPS state, the ground state of the extended object \cite{Bergshoeff:1997kr,Bandos:2001jx,Bandos:2002kk}, while another half of supersymmetry is spontaneously broken.
Namely, due to the worldvolume reparametrization invariance and local kappa-symmetry, the propagating fields on the membrane worldvolume are a scalar $\varphi(\xi)$ associated with membrane fluctuations in the transverse direction of four-dimensional space-time  (e.g. $\varphi=x^3(\xi)$) and two of four fermionic fields $\theta(\xi)$ and $\bar\theta(\xi)$. These fields form an $\mathcal N=1$, $d=3$ Goldstone supermultiplet associated with a half of $\mathcal N=1$, $D=4$ supersymmetry spontaneously broken by the presence of the membrane. The broken supersymmetry is non-linearly realized on the Goldstone supermultiplet, and the membrane action describes its coupling to the special chiral superfield $S$.

If we set $k=0$, the above action reduces to that describing a membrane moving in empty flat $\mathcal N=1$, $D=4$ superspace  \cite{Bergshoeff:1987cm,Achucarro:1988qb,Abraham:1990nz}.
 The action takes the form
\be\label{susym0}
S_{\rm free}=-T_0\int_{\calm_3}\d^3\xi\sqrt{-\det h} -\frac 1{4\pi} \left(\bar c\int_{{\calm_3}} {\mathcal C^0_3}+c.c\right),
\ee
where  $T_0=\frac {|c|}{4\pi}$ is the free membrane tension, while $\frac c{4\pi}=|T_0|e^{i\arg\, c}$ is associated with the membrane tensorial `central' charge in the $\mathcal N=1$, $D=4$ superalgebra generated by conserved supercharges which can be derived from this action, as in \cite{deAzcarraga:1989mza,Abraham:1990nz,Sorokin:1997ps}.

Applying the technique of \cite{deAzcarraga:1989mza}   to the supermembrane coupled to the superfield $S$ \eqref{susym1}, one gets an additional contribution of $S$ to the membrane `central' charge in the anti-commutator of the supercharges $Q_\alpha$.
This can be easily calculated when fermionic Goldstone fields are set to zero with the following result
\be
\label{MCGS}
\{Q_\alpha, Q_\beta\}
= \int \d x^m\wedge \d x^n\, \sigma_{mn\alpha\beta} \frac{\bar{c}+k\bar s}{4\pi}.
\ee
Therefore the membrane ground state preserving half of the bulk supersymmetry saturates
the BPS bound with this central charge, as we will see below.

\subsection{Membrane coupled to SYM}
\label{msym}

Let us now consider the membrane action \eqref{susym1}, where $S$ is the composite chiral superfield of the SYM multiplet \eqref{S=WW}. Now the effective membrane tension is\footnote{Note that since in the SYM case $S=\mathcal W^\alpha\mathcal W_\alpha$ is a nilpotent superfield, the presence of the non-zero constant $c$ in the membrane tension is essential. If $c$ were zero, the modulus $|S|$ of the nilpotent quantity would not be well defined. We thank Sergei Kuzenko for having emphasized this issue.}
\be\label{TM}
T_M=\frac 1{4\pi}\left|{kS}+c\right|.
\ee
It is tempting to assume that the membrane action \eqref{susym1} is associated with an effective field theory on the worldvolume of a BPS domain wall, including the explicit coupling to the SYM multiplet of its Goldstone sector associated with spontaneously broken 1/2 supersymmetry.

As a support to this assumption, let us show that, for a static membrane (i.e. setting to zero the worldvolume Goldstone fields), the action reduces to that of an $\mathcal N=1$, $d=3$ $SU(N)$ Chern-Simons theory
\be\label{susymstatic}
S_{\rm static}=-\frac{\ii k}{4\pi}\int_\calc\d^3\xi\tr\psi^\alpha\psi_\alpha
+\frac k{4\pi}\int_{\cal C}\tr \left(A\d A+\frac {2\ii}3 A^3\right)- T_0\int_\calc\d^3\xi \,,
\ee
where the $3d$ Majorana spinor $\psi_\alpha$ and the worldvolume Chern-Simons field $A_i(\xi)$ form an $\mathcal N=1$, $d=3$ supermultiplet. The action (\ref{susymstatic}), with
$T_0=0$ and $k=1$, was  obtained in \cite{Bashmakov:2018ghn} by inserting an interface operator into the SYM action.\footnote{Because of different conventions for differential forms in our paper and in \cite{Bashmakov:2018ghn}, the sign of our Chern-Simons term is opposite to that of \cite{Bashmakov:2018ghn}.}

In order to arrive at \eqref{susymstatic} starting from \eqref{susym1}, let us regard the fields $\lambda$, $A_m$ and $D$ inside $S$ as dynamical and set the worldvolume Goldstone fields to zero. In other words, we consider the membrane to be static, located at $x^3=0$ and set $x^i=\xi^i$, $\theta=\bar\theta=0$. \footnote{If, instead, we put to zero only the background fermions $\lambda$ and the worldvolume goldstini $\theta$ and $\bar\theta$, the supermembrane action \eqref{susym1} reduces to the ``axionic" membrane action of \cite{Townsend:1993wy}.}
Then the action \eqref{susym1} reduces to
\be\label{susymstatic0}
S_{\rm static}=-\frac 1{4\pi}\int_\calc\d^3\xi\left(\left|k\tr \lambda\lambda -\ c\right|-k\tr\lambda\sigma^3\bar\lambda \right)
+\frac k{4\pi}\int_{\cal C}\left[\tr \left(A\d A+\frac {2\ii}3 A^3\right)\right],
\ee
Now we may consider the equations of motion of $\theta(\xi)$ and $\bar\theta(\xi)$ \cite{Bandos:2010yy}, in which we should set all the worldvolume fields to zero. These impose the kappa-symmetry projection condition on the fermion $\chi$ \eqref{chi}
with the same sign as that in \eqref{kappaproj}
\be\label{chipro}
\chi_\alpha=-\ii\frac {ks+c}{|ks+c|}\Gamma_{\alpha\dot\alpha}\bar\chi^{\dot\alpha}=-\frac {k\tr\lambda\lambda-c}{|k\tr\lambda\lambda-c|}\sigma^3_{\alpha\dot\beta}\bar\chi^{\dot\beta}.
\ee
From the very definition of $\chi$ in \eqref{chi}, the previous condition translates into the following general constraint on $\lambda$
\be\label{chiprol}
\begin{split}
&\frac 1 2 \tr\left[F_{ij}\sigma^{ij}_\alpha{}^\beta \left(\lambda_\beta +\frac {k\tr\lambda\lambda-c}{|k\tr\lambda\lambda-c|}\sigma^3_{\alpha\dot\beta}\bar\lambda^{\dot\beta}\right)\right]
\\ &=
\tr\left[\left(\ii D \delta_\alpha{}^\beta - F_{i3}\sigma^{i3}_\alpha{}^\beta \right)\left(\lambda_\beta -\frac {k\tr\lambda\lambda-c}{|k\tr\lambda\lambda-c|}\sigma^3_{\alpha\dot\beta}\bar\lambda^{\dot\beta}\right)\right]\;.
\end{split}
\ee
If we consider a particular solution of \eqref{chiprol} such that $\lambda$ is subject to the same projection condition as $\chi$, namely\footnote{Another possible solution is to assume  that $\lambda$ is restricted by the condition with the opposite sign with respect to \eqref{lambdapro}. For this solution, from the definition of $\chi$ it will follow that on the static membrane $F_{ij}=0$. Then the Chern-Simons term trivializes and one finds that the static membrane action reduces to $\sim {\tt const}\,\int_{{\mathcal M}_3}\d^3\xi$. }
\be\label{lambdapro}
\lambda_\alpha=-\frac {k\tr\lambda\lambda-c}{|k\tr\lambda\lambda-c|}\sigma^3_{\alpha\dot\beta}\bar\lambda^{\dot\beta},
\ee
then  \eqref{chiprol} implies that on the membrane worldvolume
\be\label{F3i}
F_{3i}|_{\mathcal C_3}=0=D|_{\mathcal C_3}.
\ee
If $\lambda$ satisfies \eqref{lambdapro}, then we have
\be\label{trll}
\tr\lambda\lambda=\tr\lambda\sigma^3\bar\lambda\, e^{i\alpha},
\ee
where $\alpha=\arg (-k\tr\lambda\lambda+c)$.
From \eqref{lambdapro},  upon some algebra, we also have
\be\label{ll=l3l}
|k\tr\lambda\lambda-c|=\pm|c|-k\tr \lambda\sigma^3\bar\lambda\,.
\ee
Note that only the upper sign solution is consistent with the limit in which $\lambda\to 0$. Hence, we pick this one.

Substituting \eqref{trll} into  \eqref{ll=l3l} we find that for the plus-sign solution the arguments of $\alpha$ and $c$ are related as follows
\bea\label{c=a}
& \arg c=\alpha+2\pi n\,.&
\eea
This implies that $\alpha$ should be constant on the static membrane.

Due to the projection relation \eqref{chipro} the independent components of $\lambda_\alpha$ are
\be\label{lind}
\lambda_1=\frac 1{2} (\psi_1+\ii\psi_2),\qquad \lambda_2={e^{\ii\alpha}}\bar \lambda_1=\frac {e^{\ii\alpha}} 2 (\psi_1-\ii\psi_2),
\ee
where $\psi_\alpha=(\psi_1,\psi_2)$ is a real $SL(2,\mathbb{R})$ spinor. Hence, we finally have
\be\label{psipsi}
|k\tr\lambda\lambda-c|=|c|-k\tr\lambda\sigma^3\bar\lambda=|c|+ \frac {\ii k} 2\tr\psi^\alpha\psi_\alpha\,.
\ee
On the other hand, at $x^3=\theta=\bar\theta=0$ the membrane bosonic equations of motion reduce to
\be\label{beom}
\partial_{x^3}(|k\tr\lambda\lambda-c|-k\tr\lambda\sigma^3\bar\lambda)=k\varepsilon^{ijk3}\tr F_{ij}F_{k3}+ k\partial_i(\tr\lambda\sigma^i\bar\lambda).
\ee
The conditions \eqref{F3i} and \eqref{lind} imply that the right hand side of the above equation is zero, and taking into account \eqref{ll=l3l} we get
\be\label{beom1}
\partial_{x^3}(\tr\lambda\sigma^3\bar\lambda)|_{\mathcal C_3}=0,
\ee
that is, on the membrane worldvolume the derivative of $\tr\lambda\sigma^3\bar\lambda$ along the direction transverse to the static membrane should vanish. Note that \eqref{F3i} and \eqref{beom1} imply that the fields $A_i$ and $\lambda$ get localized on the membrane.

Finally, substituting the relation \eqref{psipsi} into the static membrane action \eqref{susymstatic0}, we get the $\mathcal N=1$, $d=3$ $SU(N)$ Chern-Simons action \eqref{susymstatic} of level $-k$. The term containing the constant tension $T_0$ completely decouples and can be removed by sending $T_0\to 0$. As we have already mentioned in the Introduction, for $k=1$ the obtained action is level/rank dual \cite{Bashmakov:2018ghn}  to the Acharya-Vafa \cite{Acharya:2001dz} worldvolume theory of the $k=1$ domain wall, but differs from the latter for $k>1$. Our action does not take into account additional worldvolume fields associated with relative fluctuations of a stack of $k$ coincident D-branes in the stringy construction of Acharya and Vafa. Nevertheless, the account of the effects of the membrane of charge $k$ (or of a stack of $k$ parallel membranes of charge 1) allows one to consistently derive the BPS domain wall tension and explicitly construct $k$-walls in the Veneziano-Yankielowicz effective theory, as we shall discuss in the next Section.

\section{SYM BPS domain walls sourced by membranes}

We shall now apply the analysis of \cite{Bandos:2018gjp} to elucidate properties of BPS domain walls in the Veneziano-Yankielowicz effective theory coupled to the membranes described by the action \eqref{susym0}. In \cite{Bandos:2018gjp} we dealt with $\mathcal N=1$, $D=4$ supergravity theories whose superpotentials experienced a jump at the position of the membrane separating two vacua. We will show that, similarly, the inclusion of the membrane in the VY theory is necessary to induce and take care of the discontinuity of the VY superpotential and the corresponding cusp of the VY potential reviewed in Section \ref{VYLa}. At the same time the contribution of the membrane tension to the overall energy density of the domain wall configuration makes it saturate the BPS bound, the missing ingredient which was sought in \cite{Kogan:1997dt}.

Since we are interested in 1/2 supersymmetric BPS domain walls interpolating between two supersymmetric vacua of the VY effective theory, we shall set the fermionic field $\chi_\alpha$ of the special chiral supermultiplet $S$ \eqref{S=WW} to zero and require that there is a residual 1/2 of $\mathcal N=1$ supersymmetry under which the variation of $ \chi$ vanishes.\footnote{Note that since we are now dealing with the  VY effective field theory and not directly with the SYM, the components of the special (three-form) chiral superfield $S$ \eqref{S=WW} are regarded  as independent space-time fields. In particular, $S$ is not nilpotent anymore.} We will also assume that the membrane which sources the VY domain walls is static. Namely, it stretches along the space-time directions $x^0,x^1$ and  $x^2$ and sits at the origin $x^3=0$ of the space-time coordinate orthogonal to the membrane.  The action describing the coupling of the scalar sector of the VY effective theory to the static membrane has the following form
\bea\label{Sb+Sm=st}
S=\int \d^4 x\,  {\mathcal L}_{\rm VY}^{\rm bos} -\frac 1{4\pi}\int \d^3\xi \left(|ks+c| {+} kC^3\right) \;, \quad
\eea
where ${\mathcal L}_{\rm VY}^{\rm bos}$ has been defined in \eqref{LagThreeFormCompVY} and \eqref{compbound}.\footnote{Though our main interest is the domain walls sourced by the membranes in the VY effective theory, the consideration of this Section is applicable to Wess-Zumino-type $\sigma$-models for a single three-form chiral superfield with a generic K\"ahler potential and superpotential. It can also be extended to several three-form chiral superfields as in \cite{Bandos:2018gjp}. }

Varying this action with respect to $s$ we find the equation of motion of the scalar field sourced by the membrane
\be\label{seq}
\Box s\, K_{s\bar s}+ \partial_m s \partial^m s\,  K_{ss\bar s}+ F\bar F  K_{s\bar s\bar s}+\bar F \bar W_{\bar s\bar s}= \frac k{8\pi} \delta (x^3) \frac {ks+c}{|ks+c|}\,,
\ee
where $F=\hat D+\ii \partial_mC^m$.

The equation of motion of the auxiliary field $\hat D$ is the same as \eqref{D}, while the three-form equation \eqref{Cm} acquires the membrane source term
%\be\label{D1}
%\partial_{S\bar S}K \hat D+\Re \partial_SW=0\,,
%\ee
\be\label{Cm1}
\partial_{m}(K_{s\bar s} \partial_nC^n-\Im W_{s})={-}\frac k{8\pi}  \delta^3_m\delta(x^3).
\ee
The solution of \eqref{D} and \eqref{Cm1} expresses the auxiliary field $F$ as a function of the scalar field $s$, the membrane charge $k$ and the integration parameter $n$
\be\label{F=1}
F=\hat D+\ii\partial_mC^m=-\frac{16\pi^2\bar W_{\bar s}+\ii(2\pi n+2\pi k\Theta(x^3))}{16\pi^2 K_{s\bar s}},
\ee
where $\Theta(x^3)$ is the step function. The right hand side of the above equation prompts us to introduce the discontinuous superpotential
\be\label{hat W}
\hat W(s)\equiv W(s)-  \frac \ii{8\pi} (n+k\Theta(x^3))\,s\,.
\ee
It ``jumps" at the position of the membrane and thus its local minima describe two SYM vacua, one on the left of the membrane labeled  by $n$ and another one on the right labelled by $n+k$.

In addition to the above bulk field equations, we should also take into account the equation of motion of the membrane field $x^3(\xi)$, which for $\partial_ix^3=0$ reduces to
\be\label{mbe}
(\partial_3|ks+c|+k\partial_mC^m)|_{x^3=0}=0.
\ee

We are interested in 1/2 supersymmetric BPS domain wall configurations interpolating between two vacua at $x^{3}\to -\infty$ and $x^{3}\to +\infty$ separated by the membrane, i.e.
$$
\braket{s}_{-\infty}=\Lambda^3 e^{\frac{2\pi\ii n}N }\quad {\rm and }\quad
\braket{s}_{+\infty}=\Lambda^3 e^{\frac{2\pi\ii (n+k)}N }\,.
$$
According to general properties of such domain walls (see e.g. \cite{Abraham:1990nz,Dvali:1996xe,Shifman:2009zz}),
the domain wall profile is determined by the $x^3$-dependence of the scalar field $s(x^3)$ which is constant in the other space-time directions.
Under these assumptions the supersymmetry variation of the fermionic field $\chi$ takes the form
\be
\label{DW_susyvar}
	\delta \chi_\alpha = \sqrt{2} \ii \sigma_{\alpha\dot\alpha}^3 \bar \epsilon^{\dot \alpha} \dot s + \sqrt{2} \epsilon_\alpha\,F\,,
\ee
where $\epsilon^\alpha$ is the supersymmetry parameter and  $\dot s\equiv \frac{\partial s}{\partial x^3}$. The variation should be zero under 1/2 supersymmetry preserved by the membrane supporting the domain wall solution in question.

In Section \ref{msym} we have shown that, when a static membrane is coupled to the VY Lagrangian, half of the $\caln=1$ supersymettry is preserved provided that the fermionic field $\chi_\alpha$ satisfies the condition \eqref{chipro}, determined by the  conditions on the kappa-symmetry parameters (\ref{kappaproj}) for the static membrane configuration. The corresponding supersymmetry parameter is subjected to the same condition
\be
\label{DW_proj}
 \epsilon_\alpha=e^{\ii \alpha}	\sigma^3_{\alpha \dot \alpha} \bar\epsilon^{\dot \alpha} ,
\ee
where $\alpha$ is constant in the bulk and coincides with the argument of $ {(ks+c)}|_{x^3=0}$ on the membrane surface
\be\label{alpha}
e^{\ii\alpha}:=\frac{ks+c}{|ks+c|}\Big|_{x^3=0}.
\ee
Then, the requirement that the variation \eqref{DW_susyvar} vanishes on the domain wall solution implies that
\be
\label{DW_susyvar2}
\dot s= \ii e^{\ii\alpha} F=-\ii e^{\ii\alpha} \frac{\overline{\hat W}_{\bar s}}{K_{s\bar s}},
\ee
in which we substituted the on-shell value \eqref{F=1} of $F$.

It can be easily checked that the relation \eqref{DW_susyvar2} solves the field equation \eqref{seq}. It determines how the profile of the scalar field varies along the transverse direction for the given superpotential and K\"ahler potential.

A particular choice which makes the equations  \eqref{mbe} and \eqref{DW_susyvar2} mutually consistent is that on the membrane  $ks(0)+c$, $ks(0)$ and $c$  have the same phase $\alpha$ (modulo $2\pi$).
This is what we got by analyzing the fermionic field equations on the static membrane in Section \ref{msym} (see \eqref{c=a}).\footnote{In general, the consistency of the equations \eqref{mbe} and \eqref{DW_susyvar2} allows for different, but still related, values of the phases of $ks(0)+c$, $ks(0)$ and $c$. For simplicity we will not consider this more general situation, since in the end we will set $c=0$ anyway. The cases with $c\not =0$ should be important for studying  domain walls sourced by membranes in $\mathcal N=1$ SYM theories coupled to supergravity along the lines of \cite{Bandos:2018gjp}, but this is beyond the scope of this paper.} If we make this choice, from \eqref{DW_susyvar2} it follows that
\be\label{dReW=0}
\frac \d{\d x^3} \Re (\hat We^{-\ii\alpha})=0\,,
\ee
that is
\be\label{ReW=c}
\Re  (\hat We^{-\ii\alpha})={\tt const}
\ee
at each point along $x^3$ including the position of the membrane ($x^3=0$).

We are now ready to compute the energy density (namely, the tension) of the domain wall configuration sourced by the membrane. It is determined by the on-shell value of the action \eqref{Sb+Sm=st}
 \be
\label{DW_T}
S_{\text{on-shell}} \equiv -\int \d^3\xi  \;  T_{\rm DW}\,, \quad d^3\xi:=\d x^0\wedge \d x^1\wedge \d x^2\,.
\ee
 Substituting into \eqref{Sb+Sm=st} the solution (\ref{F=1}) of the auxiliary field equations, assuming that $s$ only depends on $x^3$ and taking into account the form of the boundary term \eqref{compbound} we get
\be
\label{DW_Sa1}
S = -\int \d^3 \xi\,\d x^3\,  \left(K_{s\bar s}  \dot s \dot{\bar s}+\frac 1 {K_{s\bar s}} \hat{ W}_s \bar{\hat W}_{\bar s}  \right) -  \int \d^3 \xi\,\d x^3\, \, \delta(x^3)\, T_{\rm M}\; ,
\ee
where
\be
\label{TM:=}
T_{\rm M} = \frac{|ks+c|}{4\pi}
\ee
is the membrane tension.
The action (\ref{DW_Sa1}) can be more elegantly written in the \emph{BPS}-form
\be
\label{DW_Sb1}
\begin{split}
	S = \int \d^3 \xi\,\d x^3\,  \Bigg[&-K_{s\bar s} \left(\dot s \pm \ii e^{\ii \delta}\bar{\hat W}_{\bar s}/K_{s\bar s}\right) \left( \dot{\bar s}\mp \ii e^{-\ii \delta}   \hat{W}_s/K_{s\bar s} \right)
	\\
	&\mp \ii \left(\dot s\, \hat{W}_s e^{-\ii \delta} - \dot{\bar s}\, \bar{\hat W}_{\bar s} e^{\ii \delta} \right)  \Bigg]-  \int \d^3 \xi\,\d x^3\, \, \delta(x^3)\, T_{\rm M}
\end{split}
\ee
where $\delta$ is an arbitrary phase.

If we take $\delta=\alpha$, then the first term of \eqref{DW_Sb1} vanishes due to \eqref{DW_susyvar2} for the upper sign,  and we get the following on-shell value of the action
\be
\label{DW_Sd1}
	S = \int \d^3 \xi\,dx^3\,2\Im\left(\dot{s}\, \hat{W}_s e^{-\ii \alpha}  \right) -  \int \d^3 \xi\,\d x^3\, \, \delta(z)\, T_{\rm M}\,.
\ee
Now the integration along the transverse direction may be easily performed by noticing that, due to the form \eqref{hat W} of $\hat W$,
\be\label{d3W=}
\dot{s}\,\hat{W}_s =\frac{\d}{\d x^3} \hat{W}  +\frac {\ii k}{8\pi}  s \,\delta(x^3)
\ee
and we arrive at
\be
\label{DW_Se}
\begin{split}
	S &=  -\int \d^3 \xi\,\left(T_{\rm M}-\frac 1{4\pi}\Re (ks(0) e^{-\ii\alpha})\right)- \int \d^3 \xi\,2\,\Im [(\hat{W}_{+\infty}-\hat{W}_{-\infty})e^{-\ii(\alpha-\pi)}]\,.
\end{split}
\ee
In view of \eqref{ReW=c} and requiring the non-positive definiteness of the second term of \eqref{DW_Se}, we find that the phase of $\hat{W}_{+\infty}-\hat{W}_{-\infty}$ coincides with $\alpha-\frac{\pi}2\,(\!\!\mod{2\pi})$, and
(remembering that on the membrane  $\arg (k s(0))=\arg c=\alpha$) we see that the energy per unit area of this system is
\be\label{T}
T=2\,|\hat{W}_{+\infty}-\hat{W}_{-\infty}|+\frac{\left|c\right|}{4\pi}.
\ee
The first term of this expression
\be\label{TDW}
T_{\rm DW}=2\,|\hat{W}_{+\infty}-\hat{W}_{-\infty}|
\ee
is the tension of the domain walls  saturating the BPS bound  (see, for example, \cite{Abraham:1990nz,Dvali:1996xe,Shifman:2009zz}). The second term is the contribution of the free membrane of tension $T_0=\frac{|c|}{4\pi}$.
For $T_0=0$, the contribution of the membrane tension $T_M$ completely cancels the `jump' $|ks(0)|/4\pi$ of the superpotential along $x^3$ in \eqref{DW_Se}, and \eqref{T} reduces to \eqref{TDW}. We will then set $c=0$ by now. As we have already mentioned, the membranes with non-zero $c$ should play a role in studying supergravity domain walls (see \cite{Bandos:2018gjp} for a review and references).

On the other hand, if the membrane were not present, i.e. $T_M=0$, and the superpotential is discontinuous at $x^3=0$ then from \eqref{DW_Se} we would get the tension
\be\label{T<}
T_s=T_{\rm DW}-\frac{|ks(0)|}{4\pi}
\ee
whose value is less than that of the BPS bound, but still cannot be negative since this contribution comes from the quadratic terms of the VY part of the action \eqref{DW_Sa1}. This discrepancy was found in \cite{Kogan:1997dt} for the Veneziano-Yankielowicz superpotential \eqref{VYsp} interpolating between two vacua (as in \eqref{hat W}). In \cite{Kogan:1997dt} it was suggested that at the cusp of the VY potential there should leave an object, associated with integrated heavy modes of the theory whose tension compensates the above negative contribution and restores the BPS value of the domain wall tension. As we have just shown, this object is the dynamical membrane, described by the action \eqref{susym1}, which sources the domain wall solutions.
We have thus shown that the tension of the BPS domain-wall+membrane configurations interpolating between two supersymmetric vacua \eqref{gcond} in the VY effective theory coincides with the value of the tension of the BPS domain walls in $\mathcal N=1$ SYM \cite{Dvali:1996xe}, i.e.
\be\label{TDWsym}
T^{{}^{\rm SYM}}_{\rm DW}=\frac {N\Lambda^3}{8\pi^2}\left|e^{2\pi\ii\frac{n+k}N}-e^{2\pi\ii\frac{n}N}\right|=\frac {N\Lambda^3}{4\pi^2}\,\left|\sin{\frac{\pi k}N}\right|\,.
\ee
Let us also remind that the membrane's own tension is
$$
T_M=\frac{|ks(0)|}{4\pi}.
$$

This result is in agreement (for $c=0$) with the calculation (based on the techniques of e.g. \cite{deAzcarraga:1989mza,Cvetic:1992sf,Cvetic:1992st})  of the total tensorial central charge of the $\mathcal N=1$, $D=4$ superalgebra generated by  the domain-wall+membrane system
\be\label{totalcc}
\{Q_\alpha, Q_\beta\}
= \int \d x^m\wedge \d x^n\, \sigma_{mn\alpha\beta}\left[2 \ii (\overline{W}_{+\infty}-\overline{W}_{-\infty})+\frac{\bar{c}}{4\pi}\right].
\ee

\subsection{Multiple membranes and \texorpdfstring{$k$}{k}-walls}\label{multi}
Let us now consider the case in which instead of the single membrane of charge $k$ we have $k$ parallel membranes of charge $1$ located at different points along $x^3$. We will show that when all these membranes preserve the same 1/2 supersymmetry, i.e. when the phase of the field $s$ is the same on all the membranes,
the overall tension of the domain $k$-wall configuration created by these branes is equal to the domain-wall tension \eqref{TDWsym} sourced by the single membrane of charge $k$. In the same way and with the same result one could consider the case of several membranes of different charges $k_I$  whose sum is equal to $k$, but we will not do it to make the presentation simpler. Because of the same reason, in what follows, we will also set the bare tension $T_0=|c|$ of the membranes to zero.

Let us assume that the $k$ parallel static membranes of the three-form charge 1 are situated at the points $y_I$ ($I=1,...,k$) along the transverse direction $x^3$. Then the action \eqref{Sb+Sm=st} gets modified as follows
\bea\label{Sb+Sm=st-m}
S=\int \d^4 x\,  {\mathcal L}_{\rm VY}^{\rm bos} -\frac 1{4\pi}\sum_{I=1}^{k}\int \d^4x \,\delta(x^3-y_I) \left(|s| +C^3\right) \;. \quad
\eea
The solution of the equations of motion of the fields $\hat D$ and $C_3$ results in the following ``jumping" superpotential
\be\label{hat Wm}
\hat W(s)= W(s)-\frac \ii{8\pi}\left(n+\sum\limits_{I=1}^k\Theta(x^3-y_I)\right)s\; ,
\ee
and
\be\label{F=k}
F=\hat D+\ii\partial_mC^m=-\frac{\overline{\hat W}_{\bar s}}{K_{s\bar s}}\; .
\ee
Moreover, the $s$-field equation of motion takes the form
\be\label{seq-k}
\Box s\, K_{s\bar s}+ \partial_m s \partial^m s\,  K_{ss\bar s}+ F\bar F  K_{s\bar s\bar s}+\bar F \bar W_{\bar s\bar s}= \frac 1{8\pi} \sum_{I=1}^k\delta (x^3-y_I) \frac {s}{|s|}\,.
\ee
One can see that this equation is consistent with the 1/2 BPS equation \eqref{DW_susyvar2} if the values of the phases of the field $s(y_I)$ on each of the membrane are the same. Otherwise, generically, the supersymmetry would be completely broken and no BPS domain walls would form.

As in the previous Section, the calculation of the tension of the domain wall configuration sourced by the $k$ parallel membranes and interpolating between the $n$-th and $(n+k)$-th vacuum gives
\bea
\label{TDW=kmem}
	T_{\rm DW} =   2 \Im{\left((\hat W_{\infty} -\hat W_{-\infty}) e^{-\ii(\alpha-\pi)} \right)} +
\frac 1{4\pi}	\sum\limits_i \left(|s(y_I)| -\Re{(s(y_I)e^{-\ii\alpha})}\right) \; . \qquad \eea
Since the phases of $s(y_I)$ are equal to $e^{\ii\alpha}$, the terms under the sum in \eqref{TDW=kmem} cancel each other, and taking into account \eqref{TDWsym} we again get the correct tension of the BPS $k$-wall.
However, though the above general consideration points at a possible existence of BPS domain walls sourced by multiple separated membranes, as we will see, these are not realized (as regular solutions) in the case of the Veneziano-Yankielowicz model.

\section{BPS domain-wall solutions in the Veneziano-Yankielovicz effective theory}

Let us now analyze solutions of the BPS equation \eqref{DW_susyvar2} describing  domain $k$-walls in the VY theory. Remembering that the VY K\"ahler potential $K$ and superpotential $W$ have, respectively, the form \eqref{K=} and \eqref{VYsp} and  $\hat W$ was defined in \eqref{hat W} we rewrite the equation \eqref{DW_susyvar2} in the following form
\be\label{scalarflowVY}
\dot s={9}\ii\rho N(s\bar s)^{\frac 23} e^{\ii\alpha}\left({\ln\frac{\Lambda^3}{|s|}+\ii \arg s}-\frac{2\pi \ii}N( n+k\Theta(x^3))\right), \quad \alpha =\arg (ks(0)).
\ee
In addition we should take into account \eqref{ReW=c} which is the consequence of \eqref{DW_susyvar2}. For the case under consideration it follows from \eqref{scalarflowVY} and takes the form
\be\label{ReW=cVY}
\Re \left[e^{-\ii\alpha} s\left(\ln \frac{\Lambda^{3N}}{|s|^N}+N+\ii(2\pi n+2\pi k \Theta(x^3)-N\arg s)\right)\right]=C.
\ee
The values of the constants $\alpha=\arg (ks(0))$ and $C$ are found by imposing the asymptotic conditions describing domain wall solutions interpolating between the $n$-th and the $(n+k)$-th SYM vacuum, i.e.
$\overline {\hat W}_{\bar s}\vert_{x^3=\pm \infty}=\dot s\vert_{x^3=\pm \infty}=0$ and
$$
s_{-\infty}=\Lambda^3 e^{2\pi i \frac nN}, \qquad s_{+\infty}=\Lambda^3 e^{2\pi i \frac {n+k}N}\,.
$$
We thus get
$$
\cos \left(\frac{2\pi(n+k)}N-\alpha\right) =\cos \left(\frac{2\pi n}N-\alpha\right)
$$
and hence
\be\label{alphavalue}
\arg (ks(0))=\alpha=\pi m + \frac {\pi (2n+k)}{N}, \quad  m \in {\mathbf Z}.
\ee
and
\be\label{Deltaform}
C={N\Lambda^3 }\,\Re \,e^{\ii\left(2\pi  \frac {n}N-\alpha\right)}=(-)^{m} N\Lambda^3 \cos{ \frac{\pi k}N}.
\ee
Let us now remember that, as follows from \eqref{DW_Se}, the quantity $2\Im (\hat W_{+\infty}-\hat W_{-\infty})e^{-\ii(\alpha-\pi)}$ must be non-negative. This imposes the following condition on $m$ in \eqref{alphavalue}
\be\label{m=}
(-)^{m}\sin\frac{\pi k}N\geq 0.
\ee
So, we should take
\be\label{m=2l-1}
m=2l\quad {\rm for}\quad 0<k< N
\ee
and
\be\label{m=2l}
m=2l+1\quad {\rm for}\quad -N<k<0,\quad l \in {\mathbb Z}.
\ee

Substituting \eqref{alphavalue} and \eqref{Deltaform}  into \eqref{ReW=cVY} we get
\be\label{ReW=cVY2}
|s| \,\left(\ln \frac{\Lambda^{3}}{|s|}+1\right)\cos\left(\beta-\frac {\pi k}{N}\right)-|s|\,\sin\left(\beta-\frac {\pi k}{N}\right)\,
\left(\frac{2\pi k}N \Theta(x^3)-\beta\right) = \Lambda^3 \cos{ \frac{\pi k}N},
\ee
where $\beta(x^3)\equiv\arg s-\frac{2\pi n}N$. In view of the relation \eqref{alphavalue} and the conditions imposed on $m$ by \eqref{m=} for $0<|k|<N$ we have
\be\label{boundaryb+}
\beta|_{-\infty}=0, \qquad \beta(0)=\frac {\pi k}{N}+2\pi l, \qquad \beta|_{+\infty}=\frac{2\pi k}N \,.
\ee
We will see that for having continuous domain wall configurations the natural choice of the value of $l$ in $\beta(0)$ is $l=0$.

For $|k|=N$ (the case in which the membrane separates the same vacuum and the overall tension is zero) eqs. \eqref{ReW=cVY} and \eqref{m=} do not impose any restriction on the value of $\alpha$, and
 $C$ depends on $\alpha$ as follows
\be\label{CN}
C_N=N\Lambda^3\cos \left(\frac {2\pi n}N-\alpha\right)\, .
\ee
The $|k|=N$ case will be considered in Section \ref{k=N}.

Using eqs. \eqref{alphavalue}, \eqref{m=2l-1}, \eqref{m=2l} and \eqref{ReW=cVY2} one can show that the complex equation \eqref{scalarflowVY} reduces to the following independent real ordinary differential equation for  $0<|k|<N$
\bea\label{BPSF2}
\frac k{|k|}\frac{\dot\beta}{9\rho N\Lambda}=- \left({\frac{|s|}{\Lambda^3}}\right)^{\frac 13} \cos\left(\beta-\frac {\pi k}N\right)+\,\left({\frac{|s|}{\Lambda^3}}\right)^{-\frac 23}\, {\cos{ \frac{\pi k}N}}\,.
\eea
We will now study the solutions of the equations \eqref{ReW=cVY2} and \eqref{BPSF2} for different values of $k$ and $N$. As one can see, looking at the left hand side of \eqref{BPSF2}, the characteristic thickness of all the walls is of order $(\rho N\Lambda)^{-1}$ and decreases at large $N$.  If the parameter $\rho$ of the K\"ahler potential also depends on $N$ as $\sim \frac 1N$, the width of the domain walls will not vary with $N$.

\subsection{\texorpdfstring{$\frac {\pi |k|}N<1$}{π|k|/N<1} and \texorpdfstring{$|k|=\frac N3$}{|k|=N/3}}\label{largeN}
For closely situated vacua, i.e. when  $\frac {\pi |k|}N< 1$   on the \emph{each side} of the membrane,  the BPS equations can be solved perturbatively by expanding them in powers of infinitesimal  $\frac{\pi k}N$, $\beta$ (or $\beta-\frac {\pi k}N$), and $\frac{\delta |s|}{\Lambda^3}=\frac{ |s|}{\Lambda^3}-1$  which are of the same order. In these cases we should naturally set $l=0$ in eq. \eqref{boundaryb+}.

\begin{figure}[h!t]
    \centering
	\includegraphics[width=8cm]{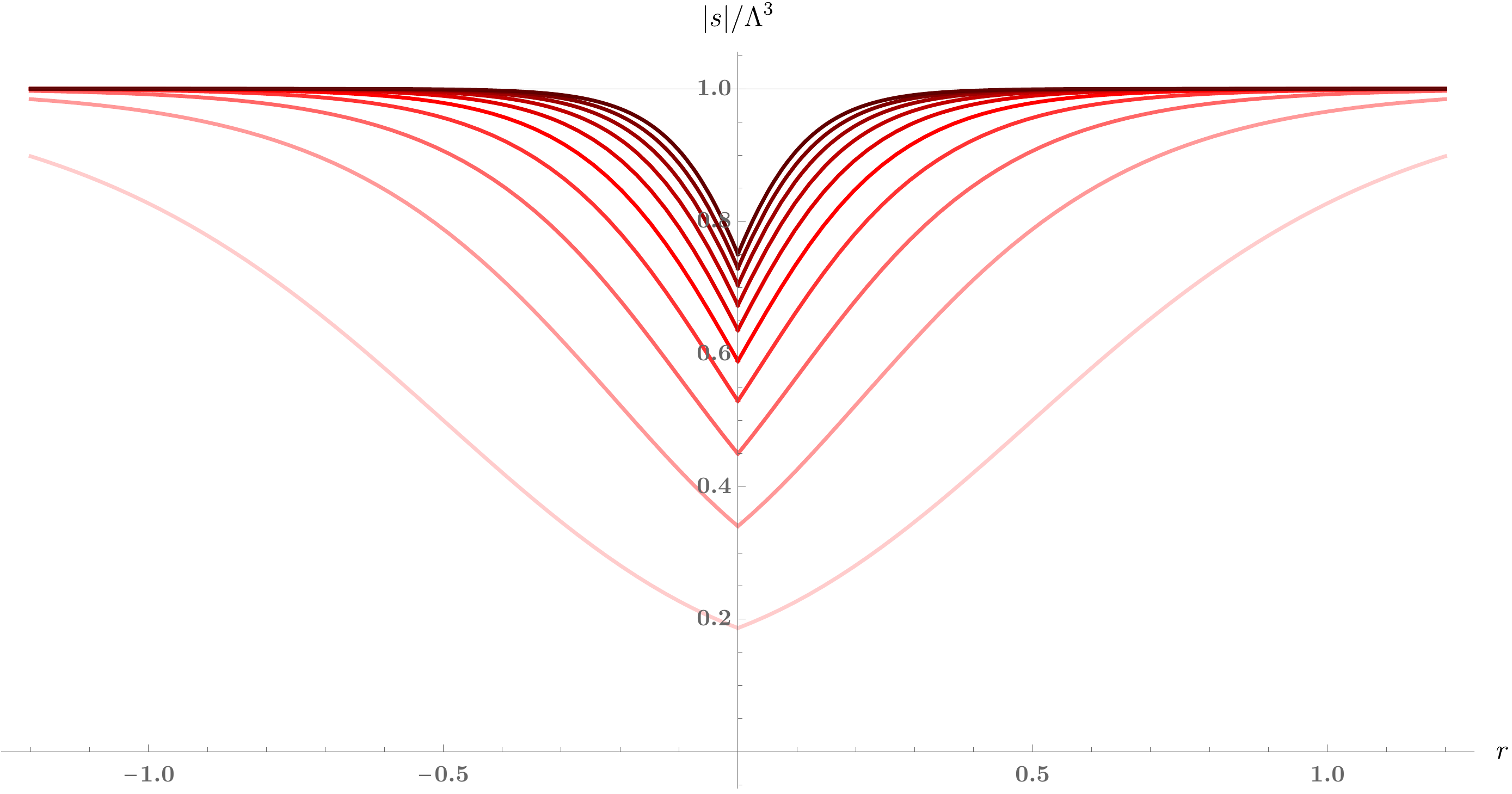} \includegraphics[width=8cm]{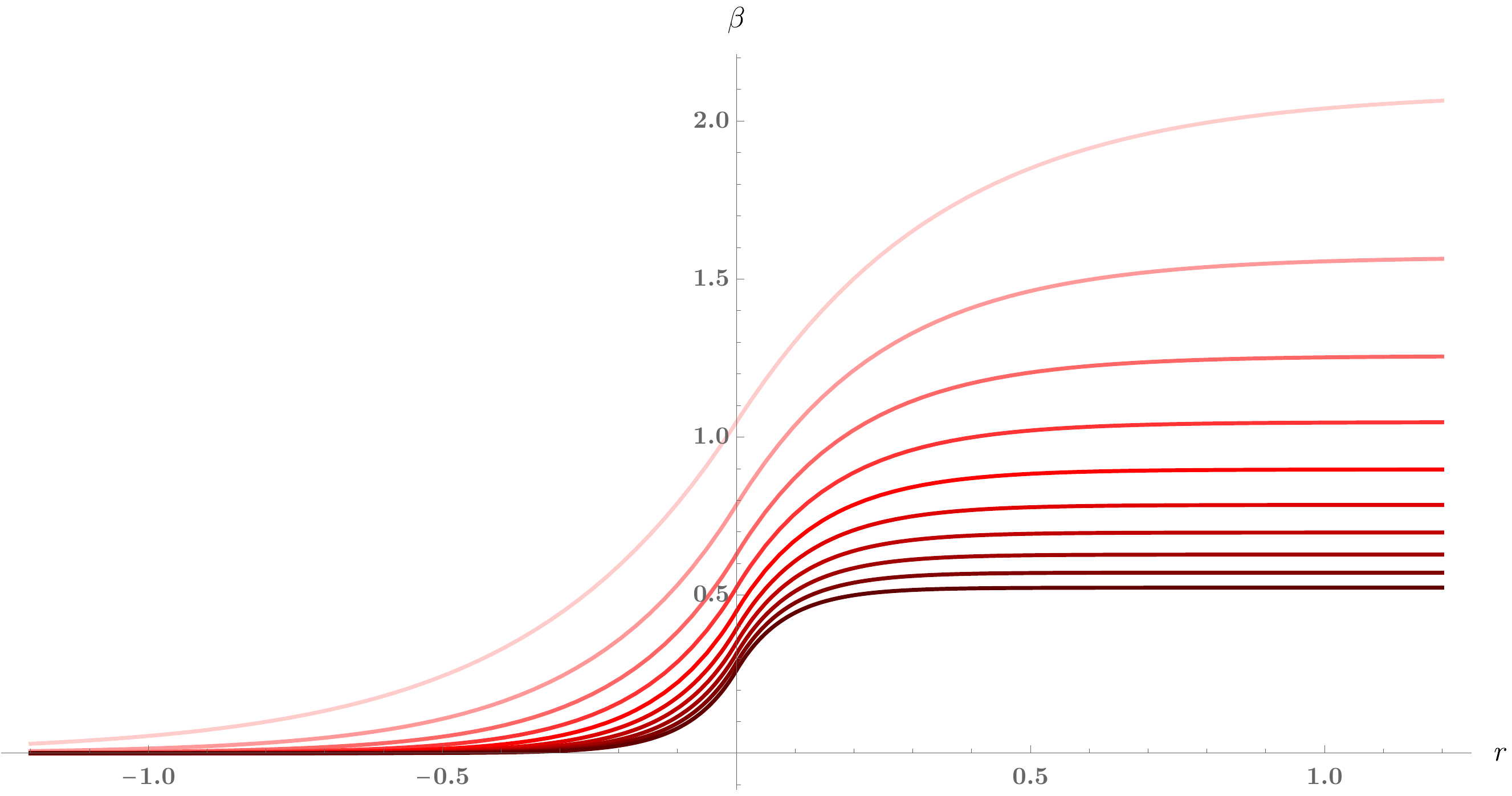}

    \caption{\footnotesize Flow of $\frac{|s|}{\Lambda^3}$ (on the left) and the phase $\beta(x^3)$ (on the right) along $r=9\rho \Lambda x^3$ for different values of $N$, with fixed $k=1$. $N$ are chosen in the interval $[3,12]$, with darker colors corresponding to larger $N$ (alternatively, one might keep $N$ fixed and vary $k$). $\frac{|s|}{\Lambda^3}$ takes the vacuum value $1$ at $x^3=\pm\infty$, decreases and has a cusp at $x^3=0$ where the membrane is sitting.  The flow of $\beta$, starts form $\beta_{-\infty} = 0$ on the left, passes through $\beta(0) = \frac{\pi}{N}$ on the membrane and reaches $\beta_{+\infty} = \frac{2\pi}{N}$ on the right. Thickness of the domain wall solutions decreases when $N$ increases. This can be fixed by choosing $\rho=\frac 1N$ in the K\"ahler potential.}\label{fig:LargeN_flows}
\end{figure}

\begin{figure}[h!t]
    \centering
	\includegraphics[width=8cm]{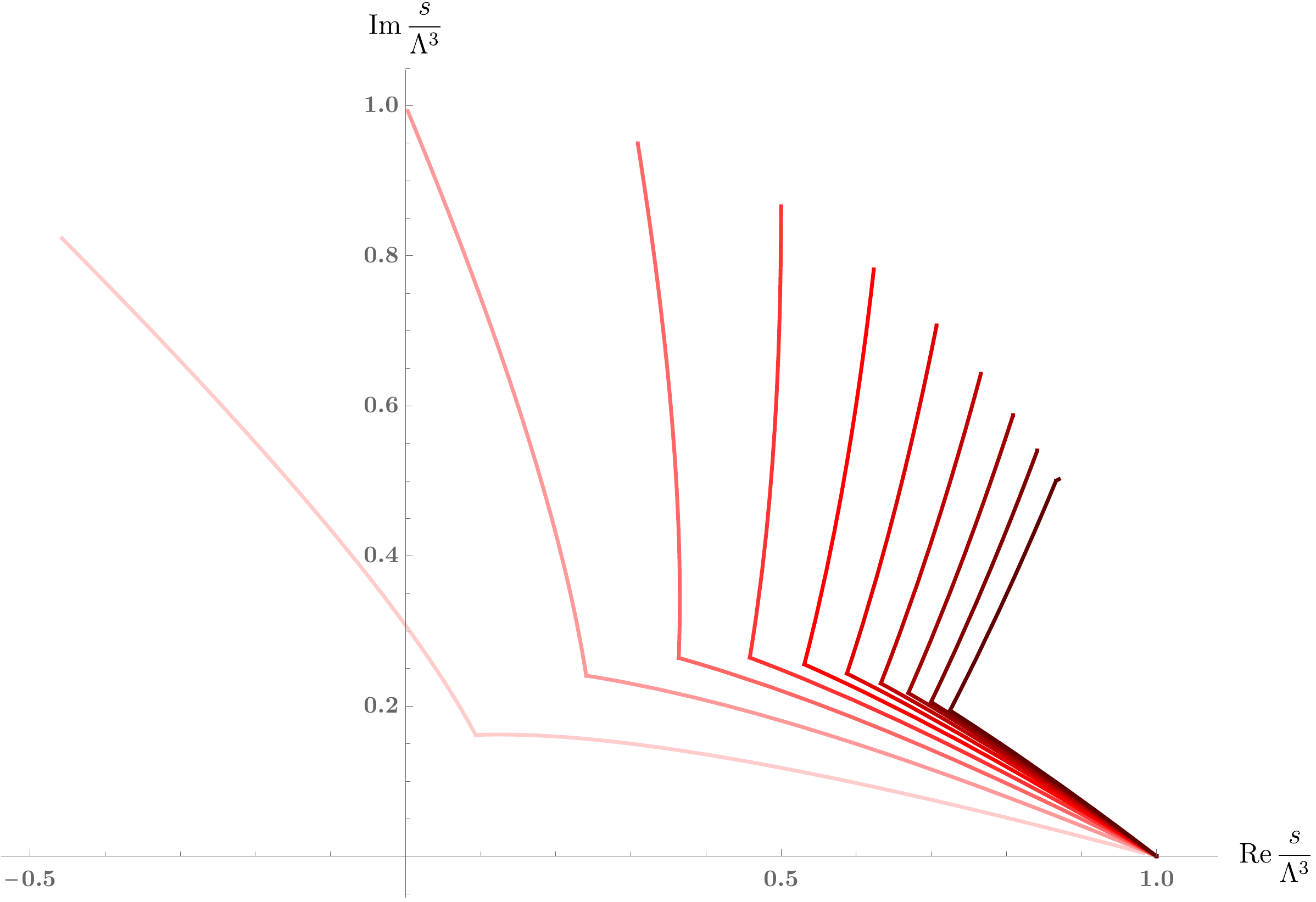}
	 \caption{\footnotesize{Behaviour of $s$ along $x^3 \in(-\infty,+\infty)$ in the complex plane (for $k=1$ and $N$ varying from 3 to 12).  Darker colors correspond to larger $N$. At the point where the membrane is located, $s(x^3)$ has  a cusp.} }\label{fig:LargeN_s}
\end{figure}

\begin{figure}[H]
    \centering
	\includegraphics[width=8cm]{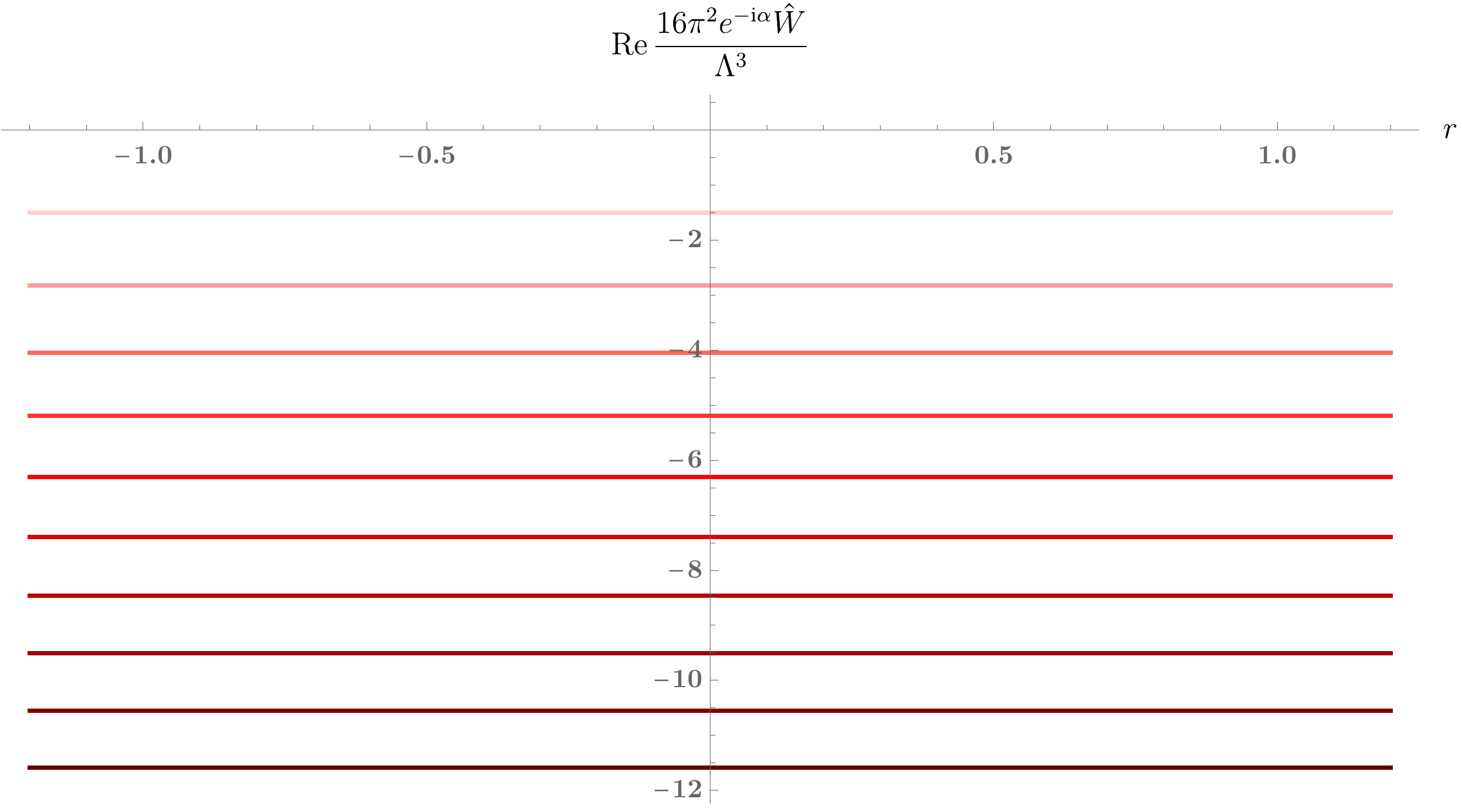} \includegraphics[width=8cm]{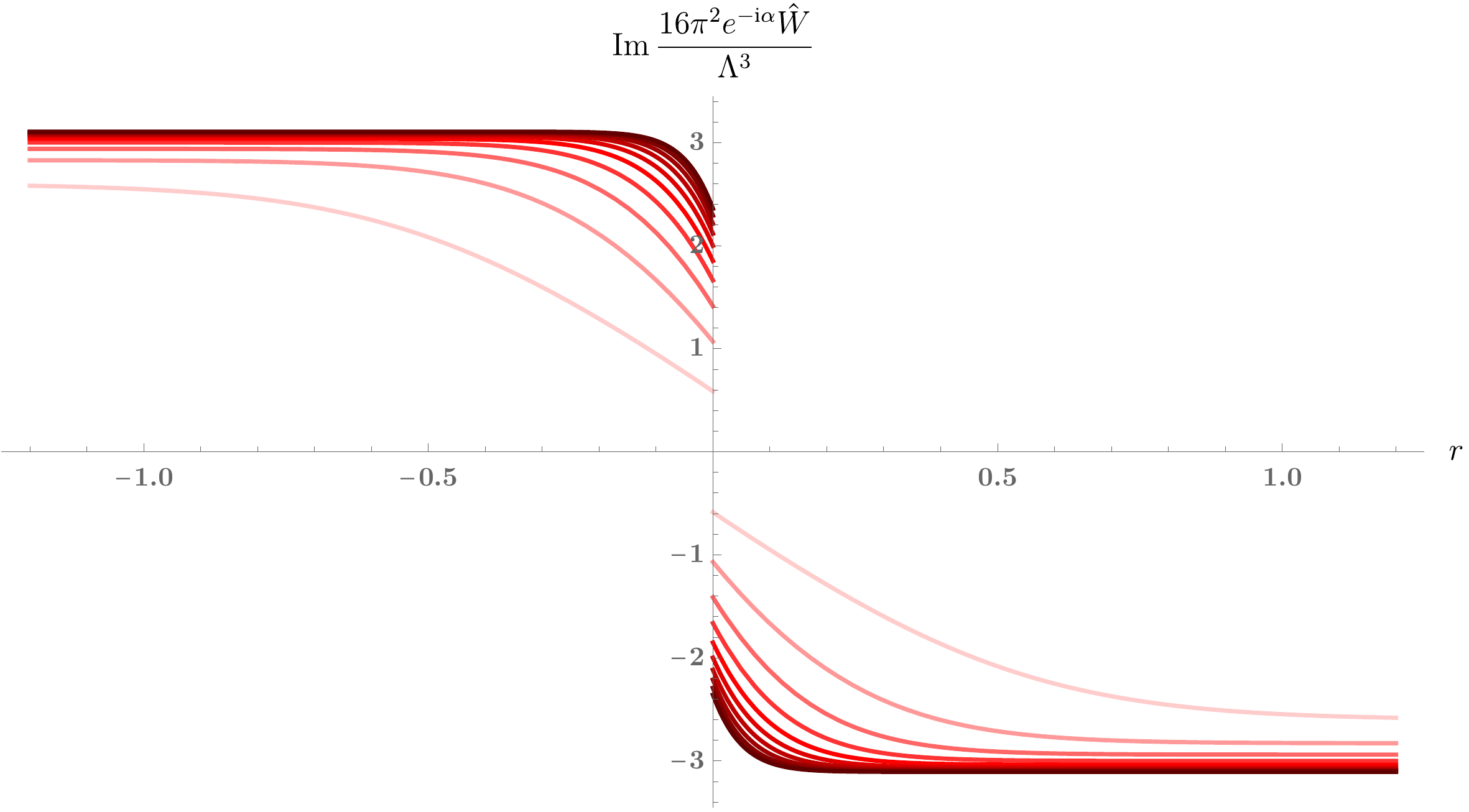}
	 \caption{\footnotesize{Behaviour (for $k=1$ and $N$ varying from 3 to 12) of the real and imaginary part of $\frac{16\pi^2\hat W e^{-\ii\alpha }}{\Lambda^3}$ ($\alpha=\frac {\pi(1+2n)}N$) along $r=9\rho \Lambda x^3$.  Darker colors correspond to larger $N$. The `jump' of $\Im\frac{16\pi^2\hat W e^{-\ii\alpha }}{\Lambda^3}$ depicted on the right is proportional to the membrane tension \eqref{TM:=} with $c=0$.} }\label{fig:LargeN_ReW_ImW}
\end{figure}

To be concrete, let us consider the case $k>0$. Then (if we require that $s(x^3)$ is continuous through the membrane), to the second order in $\beta$ the BPS equations simplify to
\be\label{largeN0}
\frac{\delta|s|}{\Lambda^3}=\pm\left(\beta+2\Theta(x^3)(\frac {\pi k}N-\beta)\right)\,
\ee
and
\be\label{dotbpert}
\frac{\dot\beta}{9\rho N\Lambda}=-\frac{\delta|s|}{\Lambda^3}+\frac 16 \left(\frac{\delta|s|}{\Lambda^3}\right)^2-\left(\beta-\frac {\pi k}N\right)\left(\frac {2\pi k}N \Theta(x^3)-\beta\right).
\ee
The exact  solution for these equations which is consistent with the boundary conditions \eqref{boundaryb+} exists for the choice of the lower sign in \eqref{largeN0}. For $x^3<0$ we get
\be\label{p1}
-\frac{\delta|s|}{\Lambda^3} =\beta=\frac {\pi k}N\,\left(1-\frac {\pi k}N\right)\left(\left(1+\frac 16\frac {\pi k}N\right) e^{-9\rho\Lambda { N}(1-\frac {\pi k}N)x^3}-\frac 76\frac {\pi k}N\right)^{-1}
\ee
and for $x^3\geq 0$
\be\label{p2}
\frac{\delta|s|}{\Lambda^3}+\frac {2\pi k}N=\beta=\frac {2\pi k}N-\frac {\pi k}N\left(1-\frac {\pi k}N\right)\left(\left(1+\frac 16\frac {\pi k}N\right)e^{9\rho\Lambda { N}(1-\frac {\pi k}N)x^3}-\frac 76\frac {\pi k}N\right)^{-1}.
\ee
Up to the second order in $\frac {\pi k}N$ the above solutions take the following form
$$
\beta=\Theta(x^3) \frac {2\pi k}N-\frac {x^3}{|x^3|} \frac {\pi k}N e^{-9\rho\Lambda { N}(1-\frac {\pi k}N)|x^3|}
\left(1-\frac 7 6\frac {\pi k}N  \left(1-
e^{-9\rho\Lambda { N}(1-\frac {\pi k}N)|x^3|}
\right)\right).
$$
The perturbative solutions are in agreement with the corresponding numerical solutions of the full BPS equations \eqref{ReW=cVY2} and \eqref{BPSF2}. This list is enlarged with the case of $|k|=\frac N3$ for which $\beta(0)=\frac {\pi |k|}N=\frac\pi 3 >1$ but is still close to unity.\footnote{Note that the cases with $k\leftrightarrow N-k$ are dual to each other since the sum of the charges of the membranes with charge $k$ and $N-k$ is $N$, i.e. equal to the periodicity of the SYM vacua. If $k\leq \frac N3$ then $N-k\geq \frac {2N}3$ and the corresponding dual configurations carry  large three-form charges, and are strongly coupled in this respect. For these configurations we have not found non-trivial (continuous) solutions of the BPS equations.}
The behaviour of the modulus $|s(x^3)|$ and the phase $\beta(x^3)$ for $|k|\leq \frac N3$ are given in Figures \ref{fig:LargeN_flows} and \ref{fig:LargeN_s}, and the behaviour of "jumping" superpotential is given in Figure \ref{fig:LargeN_ReW_ImW}.\footnote{The profiles of the found SYM BPS domain walls with $k\leq \frac N3$ are similar to those obtained in $\mathcal N=1$ $SU(N)$ super-QCD with $N_f\leq\frac N3$ (where $N_f$ is the number of flavours) in the limit  $m\to\infty$ of the mass of the flavour multiplets \cite{Smilga:2001yz}. We thank Andrei Smilga for pointing this out to us. From this perspective the membrane may be viewed as an artefact of integrated-out massive flavour modes. }

\noindent
As one can see from the plot of $\frac{|s|}{\Lambda^3}$, it tends to reach zero for smaller $N\geq 3$. As a result the solution breaks down for $N=2$, or equivalently for $k=\frac N2$. So this case should be considered separately.

\subsection{\texorpdfstring{$|k|=N/2$}{|k|=N/2}}

\noindent Physically, the cases with $k=\frac N2$ and $k=-\frac N2$ describe the same domain wall system since the difference between the two is $N$, which is the periodicity of the SYM vacua. One can also say that this domain wall system is \emph{self-dual} under $|k|\leftrightarrow N-|k|$.
In the assumption that $|s(x^3)|\not=0$, the equations \eqref{ReW=cVY2} and \eqref{BPSF2}  simplify to
\be\label{n2ln=}
\left(\ln\frac{\Lambda^3}{| s|}+1\right)=\cot{\beta}\left(\beta-\frac{k}{|k|}\pi \Theta(x^3)\right)
\ee
and
\be\label{n2dotb}
\frac {\dot\beta}{9\rho N\Lambda}=-\sin\beta\,\left({\frac{|s|}{\Lambda^3}}\right)^{\frac 13}.
\ee
Because of the minus sign on the left hand side of \eqref{n2dotb}, with the allowed choices of the $\beta(x^3)$ asymptotic  conditions \eqref{boundaryb+}, the only solution of the above equations (excluding $x^3=0$) is the step function
\be
\beta=\frac k{|k|}\, \pi\,\Theta(x^3), \qquad |s|=\Lambda^3 \,. %\qquad (x^3\not =0)
\ee
This is, obviously,  not in accord with our initial assumption (when obtaining the BPS equations) that the field $s(x^3)$ is continuous through the membrane. This indicates that the domain wall induced by the membrane with the large three-form charge \hbox{$|k|=N/2$} should be regarded as a strongly coupled system, whose internal structure is not captured by the VY effective theory.

\subsection{\texorpdfstring{$|k|=N$}{|k|=N}}\label{k=N}

In this case the vacuum on the left and the right hand side of the membrane is the same (modulo $N$) and the tension of the whole system is zero. Nevertheless, the membrane generically can break at least half of the supersymmetry. It is therefore  instructive to see if there exists a continuous  domain wall profile also in this case.
The BPS equations reduce to
\be\label{k=N11}
\frac{|s|}{\Lambda^3}\left(\ln\frac{\Lambda^3}{| s|}+1\right) =-\frac{|s|}{\Lambda^3}\tan\left(\beta-\alpha+\frac{2\pi n}N\right)\left(\beta-\frac{2\pi k}{|k|} \Theta(x^3)\right) +\frac {\cos(\alpha-\frac{2\pi n}N)}{\cos(\beta-\alpha+\frac{2\pi n}N)}
\ee
and
\bea\label{k=N22}
\frac{\dot\beta}{9\rho N\Lambda} =
-\cos\left(\beta-\alpha+\frac{2\pi n}N\right)\left({\frac{|s|}{\Lambda^3}}\right)^{\frac 13}+\,\left({\frac{|s|}{\Lambda^3}}\right)^{-\frac 23}\cos\left(\frac{2\pi n}N-\alpha\right)\,.
\eea

As we pointed out around eq. \eqref{CN}, for the case $|k|=N$ the values of $\alpha$ are not a priori restricted as in \eqref{alphavalue}. So we may try to choose an approriate value using the following reasoning. From \eqref{DW_Sa1}-\eqref{DW_Se} we know that the contribution of the VY field $ s(x)$ to the domain wall tension is non-negative
\bea\label{TN}
T_{s}&=& \int \d^3 \xi\, dx^3 \left(K_{s\bar s}  \dot s \dot{\bar s}+ \frac 1 {K_{s\bar s}} \hat{ W}_s \bar{\hat W}_{\bar s}  \right)\nonumber\\
&=&\int \d^3 \xi\,\left(2\,\Im ((\hat{W}_{+\infty}-\hat{W}_{-\infty})e^{-\ii(\alpha-\pi)})-\frac 1{4\pi}|ks(0)| \right)\geq 0.
\eea
Since in the case $|k|=N$ the first term in the second line of the above equation vanishes, the second term must be zero and hence on the membrane $|ks(0)|=0$. This can only be consistent with eq. \eqref{k=N11} if
$$\cos \,(\alpha-\frac{2\pi n}N)=0\qquad \to \qquad \alpha-\frac{2\pi n}N=\frac \pi 2+\pi m, \qquad m \in {\mathbb Z}\,. $$
With this choice of $\alpha$ the equations \eqref{k=N11} and \eqref{k=N22} reduce to
\be\label{k=N111}
\frac{|s|}{\Lambda^3}\left(\ln\frac{\Lambda^3}{| s|}+1\right) =\frac{|s|}{\Lambda^3}\cot\beta\,\left(\beta-\frac{2\pi k}{|k|} \Theta(x^3)\right)
\ee
and
\bea\label{k=N222}
\frac{\dot\beta}{9\rho N\Lambda} =(-)^{m+1}\,
\sin\beta\,\left({\frac{|s|}{\Lambda^3}}\right)^{\frac 13}\,.
\eea
As in the case $|k|=\frac N2$, the above equations do not have continuous solutions
 with the required choice of the $\beta(x^3)$ asymptotic conditions \eqref{boundaryb+}, $|s|_{\pm\infty}=\Lambda^3$ and $|s(0)|=0$.

\subsection{Multiple separated membranes do not form regular BPS domains walls in VY theory}
Suppose, as we discussed in Section \ref{multi}, that we have $k$ parallel membranes  of charge 1 distributed somehow along $x^3$.
For the 1/2 BPS configurations, the phases of the field $s(x)$ on each of them should be equal. Then, as for the single membrane of charge $k$, the properties of the BPS equations tell us that on each membrane the phase $\beta$ should be equal to $k \pi/N$ (modulo $2\pi$), and not just $\pi/N$.
This means that the flow of $\beta$ from the left of the first membrane should reach $\beta=\frac{\pi k}N$ already on this first membrane. This is, of course impossible for $k=\frac N2$, because of the sign of the right hand side of \eqref{n2dotb}.

We will now show that also for $|k|<\frac N2$,  regular solutions of the BPS equations (i.e. solutions with $\beta(x^3)$ being a continuous function through the membranes)  do not exist for parallel membranes, unless they are all concentrated at the same point of $x^3$.
As an example, let us consider two membranes with charge 1 each (i.e. $k=2$ in total). Then the algebraic equation \eqref{ReW=cVY2} takes the form
\be\label{2m}
\frac{|s|}{\Lambda^3} \,\left(\ln \frac{\Lambda^{3}}{|s|}+1\right)\cos\left(\beta-\frac {2\pi }{N}\right)-\frac{|s|}{\Lambda^3}\,\sin\left(\beta-\frac {2\pi}{N}\right)\,
\left(\frac{2\pi}N \sum_{I=1}^2 \Theta(x^3-y_I)-\beta\right) =  \cos{ \frac{2\pi}N}.
\ee
This and the differential BPS equation \eqref{BPSF2} should be solved with the boundary conditions $\beta(y_1)=\beta(y_2)=\frac{2\pi}N$, where $y_I$ are the positions of the membranes.

Between the membranes $x^3=[y_1,y_2]$ the above equation takes the form
\be\label{2m1}
\frac{|s|}{\Lambda^3}\,\left(\ln \frac{\Lambda^{3}}{|s|}+1\right)\cos\left(\beta-\frac {2\pi }{N}\right)+\frac{|s|}{\Lambda^3}\,\sin\left(\beta-\frac {2\pi}{N}\right)\,
\left(\beta-\frac{2\pi}N\right) = \cos{ \frac{2\pi}N}\,.
\ee
The derivative  $\dot\beta$ at $y_1$ and $y_2$ should have the following values
\be
\frac{\dot\beta(y_I)}{9\rho N\Lambda}=-\left(\frac{|s(y_I)|}{\Lambda^{3}}\right)^{\frac 13}+\left(\frac{|s(y_I)|}{\Lambda^{3}}\right)^{-\frac 23}\cos\frac{2\pi}N\,=\left(\frac{|s(y_I)|}{\Lambda^{3}}\right)^{\frac 13}\ln \frac{\Lambda^{3}}{|s(y_I)|}.
\ee
These are determined by the behavior of $\beta(x^3)$ from the left of the first membrane and from the right of the second membrane which should be in accordance with the plot in Fig. \ref{fig:LargeN_flows} for a single membrane of charge $k=2$ situated at $x^3=0$.

If the smooth solution for $\beta$ exists, it should have one maximum and one minimum within $[y_1,y_2]$ in which $\dot\beta=0$. These points are determined by the following relations which follow from \eqref{BPSF2} and \eqref{2m1}
\be\label{dotzero}
\frac{|s|}{\Lambda^3}=\frac{\cos{\frac{2\pi}N}}{\cos(\beta-\frac{2\pi}N)},\qquad
\ln\frac{|s|}{\Lambda^3}= (\beta-\frac{2\pi}N)\tan(\beta-\frac{2\pi}N).
\ee
Notice that these equations admit two solutions
$
\beta_{\pm}=\pm \Delta + \frac{2\pi}N,
$
where $\Delta$ should be determined by \eqref{dotzero}.

However, eq. \eqref{dotzero} also tells us that $\ln\frac{|s|}{\Lambda^3}$ should be positive (at least for small $\Delta$)
$$
\ln\frac{|s|}{\Lambda^3}>0 \rightarrow \frac{|s|}{\Lambda^3}>1.
$$
Since at $x^3=y_1$ we had $\frac{|s|}{\Lambda^3}<1$, to reach the maximum the modulus should cross the point $\frac{|s|}{\Lambda^3}=1$. The algebraic BPS equation \eqref{2m1} tells us that at this point $\beta$ should take the following value
\be
\cos\left(\beta-\frac{2\pi}N\right)+\left(\beta-\frac{2\pi}N\right)\sin\left(\beta-\frac{2\pi}N\right)=\cos\frac{2\pi}N,
\ee
which does not have solutions for $N>2$.

The above analysis can be extended to the case of an arbitrary $k\leq N/3$ with the same conclusion.

This indicates that the continuous solutions for the multiple parallel membranes do not exist. We can therefore conclude that to induce 1/2 BPS domain-wall configurations the multiple membranes of the total charge $k$ should form a  stack of coincident membranes, i.e. a (composite) membrane of charge $k$ considered in the previous Sections. Or, in other words, $k$ elementary domain walls of charge 1 should combine into a single 1/2 BPS $k$-wall, as has been asserted previously in the literature.

\section{Adding propagating massive glueballs and strings to the VY Lagrangian coupled to membranes}\label{strings}
Before concluding this paper, we would also like to briefly consider the generalization of the VY Lagrangian  in which purely gluonic bound states associated with the three-form field $C_3={}^*C_1$ in ${}^*F_4=\partial_m C^m=-\frac 14 \varepsilon^{mnpq}F_{mn}F_{pq}$ acquire a mass and become propagating degrees of freedom in addition to the gluino-balls $s(x)$ (see \cite{Farrar:1997fn} for details). This is achieved by adding to the VY Lagrangian the following mass term
\be\label{VY+U2}
\mathcal L=\mathcal L_{\rm VY}-\frac 1\delta\int \d^2\theta \d^2\bar\theta\,\frac{( U-{\bb L})^2}{(S\bar S)^{\frac 13}},
\ee
where $\delta$ is a dimensionless parameter and $U$ is the prepotential \eqref{S=D2U} determining the superfield $S$.

In the above Lagrangian we have also introduced a St\"uckelberg linear superfield~${\bb L}$, which satisfies \eqref{Lcon}, to preserve the gauge invariance \eqref{UtoU+L} of the original VY Lagrangian. Under the action of \eqref{UtoU+L},  ${\bb L}$ gets shifted by the gauge symmetry parameter~$L$
\be\label{UtoU+L1}
U'=U+ L\,, \qquad {\bb L}'={\bb L}+ L\,.
\ee
Then one can add to the Lagrangian \eqref{VY+U2} the supermembrane action \eqref{susym1}.

If the membrane has a boundary, the invariance of the membrane action under \eqref{UtoU+L1} gets broken, since its $\mathcal C_3$ term is only invariant modulo a total derivative, which in the presence of a boundary does not vanish. To restore the gauge invariance we can assume that the boundary of the membrane  $\partial {\cal M}_3= {\cal W}_2$ is the worldsheet of a string. In the case of $c=0$ the
 kappa-symmetric action of the string has the following form\footnote{
 The action \eqref{VY+U2} with  $U=0$  can be obtained as a flat superspace limit of the  superstring action coupled to ${\cal N}=1$ $4D$ supergravity and a tensor multiplet considered in \cite{Bandos:2003zk}.  A similar action  for a superstring coupled to an ${\mathcal N}=1$, $D=2+1$ supergravity via a real scalar compensator superfield $L$ was considered in \cite{Buchbinder:2017qls}.
 }
\be\label{Sstr=}
S_{\rm string}= - \frac {1} {8\pi } \int_{{\mathcal W}_2}\d^2\sigma  \sqrt{-\gamma}\,|k(U- {\bb L})| + \frac k {4\pi } \int_{{\mathcal W}_2} B_2 \; .  \qquad
\ee
%\det{\gamma_{\mu\nu}
Here $\sigma^{\mu}=(\sigma^0,\sigma^1)$ are worldsheet coordinates, $\g$ is the determinant of the metric $\g_{\mu\nu}$ induced on the worldsheet
$$
\g_{\mu\nu}= E^a_{\mu}E_{a\nu}\;
$$
and $B_2$ is a two form whose field strength is
\be\label{super2form}
\begin{aligned}
H_{3}=& \d B_2= \,   {\ii} E^a \wedge \d\theta^\alpha \wedge \d\bar\theta^{\dot\alpha}  \sigma_{a\alpha\dot\alpha}{\bb L} \\ & - {\frac 14}  E^b\wedge E^a \wedge  \d\theta^\alpha
\sigma_{ab\; \alpha}{}^{\beta}{D}_{\beta}{\bb L} -{\frac 14}  E^b\wedge E^a \wedge  \d \bar\theta^{\dot\alpha}
\bar\sigma_{ab}{}^{\dot\beta}{}_{\dot\alpha}\bar{D}_{\dot\beta}{\bb L}
\\&-\frac {1} {48}
  E^c \wedge E^b \wedge E^a \epsilon_{abcd} \,\bar{\sigma}{}^{d\dot{\alpha}\alpha}
  [D_\alpha, \bar{D}_{\dot\alpha}]{\bb L}
 \, .
\end{aligned}
\ee
We see that the interaction of the superstring with the Veneziano-Yankielowicz multiplet has the form similar to a Fayet-Iliopoulos term for the abelian vector supermultiplet $U$ containing the three-form dual $C_m(x)$.

The sum of the Wess--Zumino terms of the string and the membrane can be written in a manifestly gauge invariant way as
$-\frac k {4\pi }\int_{{\cal M}_3} ({\cal C}_3- \d B_2)$. This produces the contribution to the $\kappa$-symmetry variation of the sum of the membrane and the string action
$$
-\frac k {4\pi } \int_{{\cal W}_2} ( i_\kappa{\cal C}_3- i_\kappa \d B_2)\; ,
$$
where $\mathcal C_3$ was given in \eqref{super3form}.
%and $\mathcal C_3^0$ were given in \eqref{super3form} and \eqref{C03}.
This variation is cancelled by
 the kappa-symmetry variation of the Nambu-Goto term of the superstring action \eqref{Sstr=}
if the fermionic parameters of the $\kappa$--symmetry obey the conditions
\bea\label{kappa-str}
\kappa_\alpha = \, \frac {k( U-{\bb L})}{|k( U-{\bb L})|} \,  P_\alpha{}^\beta \kappa_\beta\, , \qquad
\bar\kappa_{\dot\alpha} =\, \frac {k( U-{\bb L})}{|k(U-{\bb L})|} \,  \bar P_{\dot\alpha}{}^{\dot\beta}\bar \kappa_{\dot\beta}\,
\eea
where
\bea\label{proj-str}
P_\beta{}^\alpha
&=& \frac 1 {2\sqrt{-\gamma}}\epsilon^{\mu\nu} E_\mu^a E_\nu^b \sigma_{ab}{}_\beta{}^\alpha, \nonumber\\
\bar P_{\dot\alpha}{}^{\dot\beta}&=& (P_{\alpha}{}^{\beta})^* =  - \frac 1 {2\sqrt{-\gamma}}\epsilon^{\mu\nu} E_\mu^a E_\nu^b \bar{\sigma}_{ab}{} ^{\dot\beta}{}_{\dot\alpha},\; \\
P^2&\equiv &{\mathbb I}.\nonumber
\eea

Notice that the superstring $\kappa$--symmetry conditions \eqref{kappa-str} are apparently consistent with the supermembrane projection conditions \eqref{kappaproj}.
However, the presence of the two projection conditions will generically reduce the
preserved supersymmetry by $1/4$ so that the corresponding supersymmetric solution of the interacting equations would describe $1/4$ BPS states.

In the case of generic  open supermembrane described by the action (\ref{susym1}) with $c\not=0$ the action for the superstring at its boundary is more complicated. We will describe it in Appendix \ref{AppB} together with a system of a supermembrane and superstring interacting with a complex three-form supermultiplet.

The above construction of membranes ending on strings can be used to study intersecting membranes and corresponding domain-wall junctions in SYM theories (see e.g. \cite{Shifman:2009zz} and references therein) and in $D=4$ supergravities. We hope to address these problems elsewhere.

\section{Conclusion}
We have considered the coupling of dynamical supermembranes to $\mathcal N=1$, $D=4$ $SU(N)$ super-Young-Mills and its Veneziano-Yankielowicz effective theory. The presence of the membrane spontaneously breaks half of the bulk supersymmetry. We have shown that the membrane with a three-form charge $k$ creates half-BPS domain walls interpolating between two SYM vacua (the $n$-th and $(n+k)$-th one).

One of the novel results of this paper is the explicit construction, in the Veneziano-Yankielowicz theory, of BPS domain wall configurations with the tension saturating the BPS bound \eqref{TDWsym}. These configurations consist of bulk scalar excitations of the VY theory and the membranes which source the BPS domain wall solutions of the scalar field equations. The VY superpotential is discontinuous along the wall while the effective potential of the scalar field has a cusp at the position of the membrane. As we showed, without the membranes such solutions do not exist, thus explaining and overcoming the obstructions to find pure SYM domain walls within the VY theory encountered in earlier literature. From this perspective one may regard the membranes as objects modifying the VY theory (similar to interface defects in Young-Mills theories, see. e.g. \cite{Gaiotto:2017tne}), while the entire wall intrinsically conflates both the membrane and the bulk scalar excitations of the theory.

Results of this paper can be straightforwardly extended to $\mathcal N=1$, $D=4$ super QCD theories containing matter flavours in the fundamental representation of the gauge group and can be used for studying domain walls within generalized Wess-Zumino models, such as the Taylor-Veneziano-Yankielowicz effective Lagrangian \cite{Taylor:1982bp}. One can also study less supersymmetry preserving BPS configurations, as well as domain wall junctions by introducing strings along which membranes end or intersect, as briefly discussed in Section \ref{strings}.

It would also be  instructive to  understand the relation of the membrane worldvolume action constructed in this paper (for the membrane charge $k>1$) to the $3d$ worldvolume gauge theories describing domain walls in SYM \cite{Acharya:2001dz} and SQCD \cite{Bashmakov:2018ghn}. As was discussed in the main text, for the case $k=1$ our supermembrane action (with the Goldstone fields switched off) is level-rank dual to the corresponding Acharya-Vaffa theory, while for $k>1$ we need to enlarge the action \eqref{susym1} with worldvolume degrees of freedom describing relative motion of a stack of $k$ coincident membranes.

\subsection*{Acknowledgements}
The authors are grateful to Francesco Benini, Sergio Benvenuti, Matteo Bertolini, Zohar Komargodski, Sergei Kuzenko, Luca Martucci, Andrei Smilga, Paul Townsend and Toine Van Proeyen for interest to this work and useful discussions. Work of I.B. was supported in part by the Spanish MINECO/FEDER (ERDF EU)  grants FPA 2015-66793-P and  PGC2018-095205-B-I00, by the Basque Government Grant IT-979-16, and the Basque Country University program UFI 11/55.  D.S. acknowledges support and hospitality extended to him at the ESI (Vienna) Program ``Higher spins and holography'' (March 11-22, 2019) at an intermediate stage of this work. Work of D.S. was also supported in part by the Australian Research Council project No. DP160103633. D.S. is grateful to the School of Physics and Astrophysics, University of Western Australia for hospitality during the final stage of this work.

%\newpage
\begin{appendix}
\section{Main conventions}\label{conventions}

The $D=4$ Levi-Civita symbol is
\begin{equation}
\varepsilon^{0123}=\varepsilon^{3210}=-\varepsilon_{0123}=1\,.
\end{equation}

\begin{equation}
\varepsilon_{m_1m_2m_3m_4} \varepsilon^{n_1n_2n_3n_4} = - 4! \delta^{n_1}_{[m_1} \delta^{n_2}_{m_2} \delta^{n_3}_{m_3}\delta^{n_4}_{m_4]}
\end{equation}
The $4D$ volume form is %{\color{green}[*Ig:** $\wedge$ inserted ***]}
\be
\d^4 x = \d x^3 \wedge \d x^2 \wedge \d x^1 \wedge \d x^0
\ee
and
\be
\d x^q \wedge \d x^p \wedge \d x^n \wedge \d x^m = \varepsilon^{qpnm} \d^4 x = \varepsilon^{mnpq} \d^4 x
\ee
Given a $p$-form $\omega_p$
\begin{equation}
\omega_p = \frac{1}{p!} \d x^{m_p}\,\wedge \ldots\, \wedge\d x^{m_1} \omega_{m_1\ldots m_{p}}\; ,
\end{equation}
the components of its Hodge-dual are defined as
\begin{equation}
\label{HD}
({}^*\!\omega)_{m_1\ldots m_{4-p}}=\frac{1}{p!}\varepsilon_{m_1\ldots m_{4-p} n_1\ldots n_p}\omega^{n_1\ldots n_p}
\end{equation}
For instance, the components of a three-form $C_3$ are
\begin{equation}
C_3 = \frac{1}{3!}  \d x^p\, \wedge \d x^n\, \wedge\d x^m C_{mnp}\,,
\end{equation}
and its field strength is
\begin{equation}
F_4 \equiv \d C_3\,, \qquad F_4 = \frac{1}{4!} \d x^q\, \wedge \d x^p\, \wedge\d x^n\,\wedge \d x^m F_{m n p q}  .
\end{equation}
with components
\begin{equation}
F_{mnpq} = 4\, \del_{[m} C_{n p q]}\,.
\end{equation}
The Hodge-dual of $F_4$ is
\begin{equation}
{}^*\!F_4 = \frac{1}{4!}\varepsilon^{mnpq} F_{mnpq} = \frac{1}{3!} \varepsilon^{mnpq}  \del_{[m} A_{n p q ]}\,,
\end{equation}
and
\begin{equation}
F_4 =  {}^*F_4\, \d^4 x\,.
\end{equation}

In the $3d$ worldvolume the Levi-Civita symbol is
\begin{equation}
\varepsilon^{012}=-\varepsilon^{210}= -\varepsilon_{012}=1\,,
\end{equation}
\be
\d^3 \xi = \d \xi^0 \wedge \d \xi^1 \wedge \d \xi^2
\ee
and
\be
\d \xi^i \wedge \d \xi^j \wedge \d \xi^k = \varepsilon^{ijk} \d^3 \xi
\ee
For sigma-matrices we use the conventions of \cite{Wess:1992cp}. However, we define
\be
\sigma_{mn} \equiv \sigma_{[m} \bar\sigma_{n]} = \frac12 \left(\sigma_m\bar\sigma_n-\sigma_n \bar\sigma_m\right)\,.
\ee
Finally, let us present a useful  sigma-matrix identity
\be
\left(\sigma^{mn}\sigma^{pq}\right)_\alpha{}^\beta
 =- i\varepsilon^{mnpq} \delta_\alpha^\beta- 2\eta^{m[p}\eta^{q]n} \delta_\alpha^\beta+
4\left(\sigma^{[m|[p}\eta^{q]|n]}\right)_\alpha^\beta\,.
\ee

\section{A system of a supermembrane ending on a superstring coupled to three-form and two-form supermultiplets}\label{AppB}

 In the case of the open membrane described by the action (\ref{susym1}) with $c\not=0$, to preserve $\kappa$-symmetry we need to add to the membrane action a superstring action whose $\kappa$-symmetry transformation  compensates the boundary variation of the membrane action
$$
-\frac k {4\pi } \int_{{\cal W}_2} i_\kappa{\cal C}_3- \frac 1 {4\pi } \int_{{\cal W}_2} (\bar  c i_\kappa{\cal C}_3^0 + c i_\kappa \bar{{\cal C}}_3^0)\; .
$$
The problem is that
${\cal C}_3^0$ is supersymmetry invariant only modulo a total derivative, which
one easily sees from \eqref{C03}. As such,   the superstring action should also contain corresponding contributions which are not manifestly supersymmetric.
The action takes the following form
\bea\label{Sstr=Sigma}
S_{\rm string}= - \frac 1 {8\pi} \int\limits_{{\cal W}_2}\d^2\sigma \sqrt{-\gamma}|k(U- {\bb L})+ \bar c( \theta^2 - {\bb L}_1 - i{{\bb L}}_2 )+ c( \bar  \theta^2 - {\bb L}_1+i{\bb L}_2)  | - \nonumber \\
- \frac k {4\pi } \int\limits_{{\cal M}_3} {H}_3+ \frac {\bar c} {4\pi } \int\limits_{{\cal M}_3} {\bb H}_3  +
\frac {c} {4\pi  } \int\limits_{{\cal M}_3} \bar{{\bb H}}_3 \; . \nonumber \\ {}
\eea
In \eqref{Sstr=Sigma} $ {\bb L}$,  $ {\bb L}_1$  and  ${\bb L}_2$ are  real linear superfields (i.e. satisfying \eqref{Lcon}), $H_3=\d B_2$ is the 3-form field strength constructed in terms of the real linear superfield $ {\bb L}$,
 \eqref{super2form}, while ${{\bb H}}_3$ is a complex field strength constructed as in \eqref{super2form} but with the complex combination ${\bb L}_1+i{\bb L}_2$ instead of ${\bb L}$
$$
{{\bb H}}_3= (\bar {\bb H}_3)^*= \d {\bb B}_2=H_3 \vert_{{\bb L}\mapsto {\bb L}_1+ i{\bb L}_2} \; .
$$

The sum of the action \eqref{susym1} for  the open supermmebrane  and \eqref{Sstr=Sigma} of the superstring at its end is invariant under space-time supersymmetry if the real linear superfields are transformed as follows
\be\label{susyL1L2} \delta_\epsilon {\bb L}_1 = \theta^\alpha\epsilon_\alpha + \bar\theta_{\dot\alpha}\bar{\epsilon}^{\dot\alpha}\; , \qquad \delta_\epsilon  {\bb L}_2=-i \theta^\alpha\epsilon_\alpha +i \bar\theta_{\dot\alpha}\bar{\epsilon}^{\dot\alpha}\; .  \ee
One can observe that such a transformation does not leave invariant ${{\bb H}}_3$ and $ (\bar {\bb H}_3)^*$ in
(\ref{Sstr=Sigma}), but these are compensated by the transformations of
the complex 3-form potential  ${\cal C}_3^0$ \eqref{C03}.

To understand such a non-manifest form of the supersymmetry invariance, one can consider $\Sigma^0=\theta^2$ as a particular complex linear superfield (i.e. the superfield obeying the constraint $\bar D^2\Sigma=0$). As it was discussed briefly in the footnote \ref{sugra}, $\Sigma=\Sigma^0$ can be considered as a flat-superspace limit of the gauge fixing condition (${\cal Z}=1$) for a complex linear prepotential of the chiral conformal compensator superfield
${\cal Z}= -1/4 (\bar {\cal D}^2- 8R)\bar \Sigma $ (with $( {\cal D}^2- 8\bar R)\bar \Sigma=0$)
of a special minimal complex 3-form supergravity (in the notation of \cite{Wess:1992cp}).

$S_{\rm membrane}+S_{\rm string}$ is also invariant under the local fermionic $\kappa$--symmetry \eqref{kappasymm} with the parameter obeying
\eqref{kappaproj} and (\ref{kappagamma}) on ${\cal M}_3$ and
\bea\label{kappa-str2}
\kappa_\alpha = \, \frac {k(U-{\bb L})+ \bar c(\theta^2 - {\bb L}_1- i{{\bb L}}_2 )+c(\bar\theta^2- {\bb L}_1+ i{{\bb L}}_2) }{|k(U-{\bb L})+\bar c(\theta^2 - {\bb L}_1- i{{\bb L}}_2 )+ c(\bar\theta^2- {\bb L}_1+  i{{\bb L}}_2) |} \,  P_\alpha{}^\beta \kappa_\beta\, ,
\eea
 on the worldsheet $\calw_2=\partial\calm_3$, with $P_\alpha{}^\beta= (P_{\dot\alpha}{}^{\dot\beta})^*$ defined in (\ref{proj-str}).

\subsection{String at the end of the membrane interacting with a complex three-form supermultiplet}

The action for the supermembrane  has now the following form
\be
S_{\rm mem}=-\frac 1 {4\pi} \int\limits_{{\cal M}_3}\d^3\xi \sqrt{-h}|{c}T| -\frac {\bar c} {4\pi} \int\limits_{{\cal M}_3}{\bb C}_3 -\frac {c} {4\pi} \int\limits_{{\cal M}_3}\bar{{\bb C}}_3 \;
\ee
where
\be
T=-\frac 1 4 \bar D^2 \bar \Sigma, \qquad \bar{T}= -\frac 1 4  D^2  \Sigma\; ,
\ee
are speical chiral superfields describing the complex three-form supermultiplet (see \cite{Farakos:2017jme,Bandos:2018gjp} and references therein),  $ \Sigma$ is the complex linear superfield  and
${\bb C}_3$ is similar to the three-form defined in \eqref{super3form} but in which the real superfield $U$ is replaced with $\Sigma$, $
{{\bb C}}_3={\cal C}_3 \vert_{U\mapsto \Sigma}
$.

When the membrane has a boundary, to maintain the gauge invariance and the $\kappa$-symmetry, we should extend this action with a boundary term which describes a superstring on which the membrane is ended
\bea\label{Sstr=Sigma2}
S_{\rm str}= - \frac 1 {8\pi} \int\limits_{{\cal W}_2}\d^2\sigma \sqrt{-\gamma}|\bar c(\Sigma - {\bb L}- i\tilde{{\bb L}} )+ c( \bar  \Sigma - {\bb L}+i\tilde{{\bb L}})  | +
\frac {\bar c} {4\pi } \int\limits_{{\cal M}_3} {\bb H}_3 +
\frac c {4\pi  } \int\limits_{{\cal M}_3} \bar{{\bb H}}_3 \; . \nonumber \\ {}
\eea
Here  $ {\bb L}$  and  $\tilde {\bb L}$ are two real linear superfields,  and
$$
{{\bb H}}_3= (\bar{{\bb H}}_3)^*= \d {\bb B}_2={\cal C}_3 \vert_{U\mapsto {\bb L}+ i\tilde{{\bb L}}} \; ,
$$
with ${\cal C}_3$ defined as in  \eqref{super3form} but in which $U$ is replaced with ${\bb L}+ i\tilde{{\bb L}}$.

The sum of the above actions is invariant under the $\kappa$--symmetry with the fermionic Weyl-spinor parameter obeying the constraint
\bea\label{kappa-str3}
\kappa_\alpha = \, \frac {\bar c( \Sigma - {\bb L}- i\tilde{{\bb L}})+c(\bar \Sigma - {\bb L}+ i\tilde{{\bb L}} )}{|\bar c( \Sigma - {\bb L}- i\tilde{{\bb L}})+c(\bar \Sigma - {\bb L}+ i\tilde{{\bb L}} )|} \,  P_\alpha{}^\beta \kappa_\beta\, .
\eea
where $P_\alpha{}^\beta = (P_{\dot\alpha}{}^{\dot\beta})^*$ was defined in (\ref{proj-str}).

\end{appendix}

%\maketitle  IS IGNORED %%%%%%%%%%%

%\listoftables       % ONLY IN DRAFT MODE
%\listoffigures      % ONLY IN DRAFT MODE

%%%%%%%%%%%%%%%%%%%%%%%%%%%%%%
%%%%%%%%%%%%%%%%%%%%%%%%%%%%%%

\if{}
\bibliographystyle{abe}
\bibliography{references}{}
\fi

\providecommand{\href}[2]{#2}\begingroup\raggedright\endgroup

\end{document}